\documentclass{article}

\newlength\ys
 
\usepackage{amsmath,amssymb,dsfont,paralist,amsthm,eurosym}
\usepackage[format=default,labelfont=bf]{caption}
\usepackage{placeins}
\usepackage{microtype}
\usepackage[utf8]{inputenc}
\usepackage{scalerel}
\usepackage{stmaryrd}
\usepackage{bbold}
\usepackage{fancyhdr}

\usepackage{algorithm}
\usepackage{algpseudocode}

\algnewcommand\algorithmicforeach{\textbf{for each}}
\algnewcommand\algorithmicvariables{\textbf{Variables:}}
\algnewcommand\Variables{\item[\algorithmicvariables]}
\algdef{S}[FOR]{ForEach}[1]{\algorithmicforeach\ #1\ \algorithmicdo}
\algrenewcommand\algorithmicrequire{\textbf{Input:}}
\algrenewcommand\algorithmicensure{\textbf{Output:}}
\algnewcommand\Fixedcomment[1]{\hfill\makebox[0.4\textwidth][l]{$\triangleright$ #1}}

\usepackage{graphicx}
\usepackage{tikz}
\usepackage{subcaption}
\usepackage{tikz-qtree}
\usepackage{rotating}

\usetikzlibrary{arrows,matrix}
\usetikzlibrary{matrix,calc}
\usetikzlibrary{shapes.geometric}
\usetikzlibrary{positioning}
\usetikzlibrary{calligraphy}
\usetikzlibrary{fit}

\pgfdeclarelayer{background}
\pgfsetlayers{background,main}

\usetikzlibrary{decorations.pathreplacing}
\tikzset{curlybrace/.style={decoration=brace,decorate}}
\usetikzlibrary{automata}
\usetikzlibrary{calc,positioning}
\usetikzlibrary{decorations.pathmorphing}
\usetikzlibrary{shapes}
\tikzset{trinode/.style={draw,triangle,minimum width=2.0cm}}
\tikzset{snake/.style={decorate,decoration=snake}}
\tikzset{curlybrace/.style={decoration=brace,decorate}}
\tikzset{triangle/.style={regular polygon,regular polygon sides=3}}
\tikzset{edge from parent path={(\tikzparentnode) -- (\tikzchildnode.north)}}
\usetikzlibrary{arrows.meta}

\usepackage{wrapfig}
\usepackage[rightcaption]{sidecap}
\usepackage[colorinlistoftodos]{todonotes}
\usepackage{mathtools}
\usepackage{enumitem}
\usepackage{multicol} 
\usepackage{todonotes}
\usepackage{imakeidx}         

\usepackage{hyperref}

\newtheorem{theorem}{Theorem}[section]
\newtheorem{lemma}[theorem]{Lemma}

\newtheorem{corollary}[theorem]{Corollary}
\newtheorem{observation}[theorem]{Observation}
\newtheorem{example}[theorem]{Example}
\newtheorem{definition}[theorem]{Definition}

\newtheorem{claim}[theorem]{Claim}

\newcommand{\cA}{\mathcal A}

\newcommand{\rmM}{\mathrm{M}}

\newcommand{\C}{\mathrm{C}}

\newcommand{\T}{\mathrm{T}}

\newcommand{\R}{\mathrm{R}}

\newcommand{\im}{\mathrm{im}}

\newcommand{\wt}{\mathrm{wt}}

\newcommand{\sfM}{\mathsf{M}}
\newcommand{\sfA}{\mathsf{A}}
\newcommand{\sfB}{\mathsf{B}}

\newcommand{\A}{\mathsf{A}}
\newcommand{\B}{\mathsf{B}}

\newcommand{\M}{\mathsf{M}}

\newcommand{\sfT}{\mathsf{T}}

\newcommand{\Trunc}{\mathsf{Trunc}}

\newcommand{\V}{\mathsf{V}}

\newcommand{\0}{\mathbb{0}}
\newcommand{\1}{\mathbb{1}}

\newcommand{\h}{\mathrm{h}}

\newcommand{\pos}{\mathrm{pos}}

\newcommand{\rk}{\mathrm{rk}}

\newcommand{\runsem}[1]{[\![#1]\!]^{\mathrm{run}}}
\newcommand{\initialsem}[1]{[\![#1]\!]^{\mathrm{init}}}

\newcommand{\wcl}{\mathrm{wcl}}

\newcommand{\rmST}{{\mathrm{ST}_\Sigma}}
\newcommand{\sfST}{\mathsf{ST}}
\newcommand{\rmS}{\mathrm{S}}
\newcommand{\sfS}{\mathsf{S}}

\usepackage[most]{tcolorbox}
\newtheorem{theorem-rect}[theorem]{Theorem}
\newtheorem{corollary-rect}[theorem]{Corollary}
\newtheorem{lemma-rect}[theorem]{Lemma}
\newtheorem{construction-rect}[theorem]{Construction}
\newtheorem{definition-rect}[theorem]{Definition}

\tcolorboxenvironment{theorem-rect}{
		arc=3pt,				
		colback=white,			
		enlarge top by=2mm,		
		enlarge bottom by=-1mm,	
		grow to left by=3mm,	
		grow to right by=3mm,	
		boxrule=0.1mm,			
		left=1.7mm,				
		right=1.7mm				
		}						

\tcolorboxenvironment{corollary-rect}{
		arc=3pt,
		colback=white,
		enlarge top by=2mm,
		enlarge bottom by=-1mm,
		grow to left by=3mm,
		grow to right by=3mm,
		boxrule=0.1mm,
		left=1.7mm,
		right=1.7mm
              }

\tcolorboxenvironment{lemma-rect}{
		arc=3pt,
		colback=white,
		enlarge top by=2mm,
		enlarge bottom by=-1mm,
		grow to left by=3mm,
		grow to right by=3mm,
		boxrule=0.1mm,
		left=1.7mm,
		right=1.7mm
              }

\tcolorboxenvironment{construction-rect}{
		arc=3pt,
		colback=white,
		enlarge top by=2mm,
		enlarge bottom by=-1mm,
		grow to left by=3mm,
		grow to right by=3mm,
		boxrule=0.1mm,
		left=1.7mm,
		right=1.7mm
              }

\tcolorboxenvironment{definition-rect}{
		arc=3pt,
		colback=white,
		enlarge top by=2mm,
		enlarge bottom by=-1mm,
		grow to left by=3mm,
		grow to right by=3mm,
		boxrule=0.1mm,
		left=1.7mm,
		right=1.7mm
              }

\pgfdeclaredecoration{dashsoliddouble}{initial}{
  \state{initial}[width=\pgfdecoratedinputsegmentlength]{
    \pgfmathsetlengthmacro\lw{.7pt+.5\pgflinewidth}
    \begin{pgfscope}
      \pgfpathmoveto{\pgfpoint{0pt}{\lw}}%
      \pgfpathlineto{\pgfpoint{\pgfdecoratedinputsegmentlength}{\lw}}%
      \pgfmathtruncatemacro\dashnum{%
        round((\pgfdecoratedinputsegmentlength-3pt)/6pt)
      }
      \pgfmathsetmacro\dashscale{%
        \pgfdecoratedinputsegmentlength/(\dashnum*6pt + 3pt)
      }
      \pgfmathsetlengthmacro\dashunit{3pt*\dashscale}
      \pgfsetdash{{\dashunit}{\dashunit}}{0pt}
      \pgfusepath{stroke}
      \pgfsetdash{}{0pt}
      \pgfpathmoveto{\pgfpoint{0pt}{-\lw}}%
      \pgfpathlineto{\pgfpoint{\pgfdecoratedinputsegmentlength}{-\lw}}%
      \pgfusepath{stroke}
    \end{pgfscope}
  }
}

\tikzset{small circle/.style={circle, draw=black, inner sep=0pt,outer sep=0pt, minimum size=2.5pt}}

\title{\bf \LARGE The generating power of weighted tree automata \\ with initial algebra semantics}

\date{\today\endgraf \normalsize
  \vspace{40mm}
  \hspace*{70mm}
  }

  \author{Manfred Droste\\University of Leipzig, Germany \and
    Zolt\'an F\"ul\"op\footnote{Project no TKP2021-NVA-09 has been implemented with the support provided by the Ministry of Culture and Innovation of Hungary from the National Research, Development and Innovation Fund, financed under the TKP2021-NVA funding scheme.}\\ University of Szeged, Hungary \and
    Andreja Tepav{\v c}evi{\'c}\footnote{This research was supported by the Science Fund of the Republic of Serbia, $\#$ Grant no 6565, Advanced Techniques of Mathematical Aggregation and Approximative Equations
Solving in Digital Operational Research-AT-MATADOR}\\Mathematical Institute SANU, Belgrade \& University of Novi Sad, Serbia
    \and
Heiko Vogler\\Technische Universit\"at Dresden, Germany}

\begin{document}
\maketitle

\begin{abstract}
We consider the images of the initial algebra semantics of weighted tree automata over strong bimonoids (hence also over semirings). These images are subsets of the carrier set of the underlying strong bimonoid.
We consider locally finite, weakly locally finite, and bi-locally finite strong bimonoids. We show that there exists a strong bimonoid which is weakly locally finite and not locally finite. We also show that if the ranked alphabet contains a binary symbol,
then for any finitely generated strong bimonoid, weighted tree automata
can generate, via their initial algebra semantics, all elements of the strong bimonoid.
As a consequence of these results, for weakly locally finite strong bimonoids which are not locally finite,
weighted tree automata can generate infinite images provided that the input ranked alphabet
contains at least one binary symbol. This is in sharp contrast to the setting of weighted string automata,
where each such image is known to be finite.
As a further consequence, for any finitely generated semiring, there exists
a weighted tree automaton which generates, via its run semantics, all elements of  the semiring.
\end{abstract}

\small{Keywords: strong bimonoids, semirings, weighted tree automata,  initial algebra semantics, finite image property,
universality property.}

       \section{Introduction}
       
Weighted tree automata (wta) are combinations of finite-state tree automata \cite{eng75-15,gecste84,gecste97,comdaugiljaclugtistom08} and weighted string automata (wsa) \cite{sch61,eil74,salsoi78,wec78,kuisal86,sak09,drokuivog09,drokus21}. Each wta assigns to an input tree a weight from some weight algebra; a weight algebra has a binary multiplication for accumulation of values along one run of the wta on the input tree, and a binary summation for accumulation of run values. Essentially, a wsa can be viewed as a wta over a string ranked alphabet, i.e., a ranked alphabet that consists of exactly one nullary symbol and at least one unary symbol (cf., e.g., \cite[p.~324]{fulvog09new} and \cite[Sec.~3.3]{fulvog22}).

In earlier research on wsa and wta, the multiplication of the weight algebra was assumed to distribute over summation (which is typical in semirings and fields). More recent investigations use strong bimonoids as weight algebras. A \emph{strong bimonoid} is an algebra  $\B = (B,\oplus,\otimes,\0,\1)$  where  $(B,\oplus,\0)$  is a commutative monoid,  $(B,\otimes,\1)$  is a monoid, and  $\0$ is annihilating with respect to  $\otimes$, i.e., $b \otimes \0 = 0 \otimes b = \0$  for each  $b \in B$, i.e., no distributivity is required.
For such results, we refer to \cite{drostuvog10,cirdroignvog10,drovog12} for wsa and to \cite{rad10,drofulkosvog20b,fulkosvog19,drofulkosvog21} for wta; see \cite{fulvog22} for a recent overview of the theory of wta over strong bimonoids and semirings. The class of strong bimonoids covers, e.g., the class of semirings and bounded lattices (including the large class of non-distributive bounded lattices) \cite{bir93}.
We mention that wta over even more general weight algebras in which the multiplication is generalized, were studied: wta over multioperator monoids \cite{fulstuvog12,stuvogfue09}  and wta over tree valuation monoids \cite{drogoemaemei11,droheuvog15,drofulgoe16,drofulgoe19}; for an abstract approach for wsa see \cite{gasmon18}. In this paper, we consider wta over strong bimonoids as weight algebras.

For each wsa over a semiring, its run semantics, initial algebra semantics, and its free monoid semantics are equal, cf. \cite[Ch.~VI, Cor.~6.2]{eil74} and \cite[Lm.~5]{cirdroignvog10}. This is no longer true for wsa over strong bimonoids because, e.g., matrix multiplication need not be associative. Moreover, since the set of trees $\T_\Sigma$ over a ranked alphabet $\Sigma$ does not form a monoid, the concept of free monoid semantics is not available for wta. Hence, for a wta  $\cA$, two kinds of semantics are considered: the run semantics $\runsem{\cA}$  and the initial algebra semantics $\initialsem{\cA}$. Both are mappings from the set $\T_\Sigma$  into the strong bimonoid  $\B$. For each wta over a semiring, its run semantics and its initial algebra semantics are equal, cf. \cite[Thm.~4.1]{rad10} and \cite[Lm.~4.1.13]{bor04} (see Theorem~\ref{thm:semiring-run=initial}).

Since each wsa can be viewed as a wta over a string ranked alphabet, it is interesting to compare the computational power of wta over string ranked alphabets with that of wta over arbitrary ranked alphabets, and thereby to find out which impact the branching of the input symbols has on the computational power.

In this paper, we focus on one particular aspect of the computational power of a wta, viz. its generating power. In particular, we ask whether the image of the semantics of a wta  $\cA$  is a finite set, or whether the wta  $\cA$  can generate even all elements of the given strong bimonoid as values. In trivial examples of one-state wta  $\cA$, both $\im(\runsem{\cA})$  and  $\im(\initialsem{\cA})$  are infinite.

The goal of this paper is to investigate which conditions on strong bimonoids  $\B$  and ranked alphabets  $\Sigma$  ensure for each wta  $\cA$  over  $\Sigma$  and  $\B$  that the image  $\im(\initialsem{\cA})$  of the initial algebra semantics of  $\cA$  is finite.

In the literature, several such conditions on strong bimonoids have been investigated.
A strong bimonoid $\B=(B,\oplus,\otimes,\0,\1)$  is \emph{locally finite} if for each finite subset  $A \subseteq B$, the additive and multiplicative closure of  $A$ (i.e., the strong subbimonoid generated by  $A$) is finite. It is easy to see that for each wta  $\cA$  over a locally finite strong bimonoid  $\B$  and any ranked alphabet  $\Sigma$, both  $\im(\runsem{\cA})$  and $\im(\initialsem{\cA})$  are finite (cf. \cite[Lm. 16.1.1]{fulvog22}). In fact, for the run semantics here it suffices that  $\B$  is \emph{bi-locally finite}, i.e. the two monoids  $(B,\oplus,\0)$  and  $(B,\otimes,\1)$  are locally finite. However, as examples show, bi-local finiteness of $\B$  does not suffice to guarantee the finiteness of $\im(\initialsem{\cA})$, even if  $\cA$  is a wsa, cf. \cite[Ex.~25]{drostuvog10} and \cite[proof of Thm.~5.2.5(2)]{fulvog22}. For the setting of wsa, it was shown that if the strong bimonoid  $\B$  is \emph{weakly locally finite}, then for each wsa  $\cA$  over a string alphabet and  $\B$  the image  $\im(\initialsem{\cA})$  is finite; a strong bimonoid  $\B$  is said to be weakly locally finite, if the weak closure of each finite subset  $A \subseteq B$  is finite where the weak closure of  $A$  is obtained by adding arbitrarily generated two elements and multiplying such elements with elements of  $A$  from the right. Trivially, we have the following implications for arbitrary strong bimonoids  $\B$:

\begin{center}
$\B$  locally finite \ $\Rightarrow$ \ $\B$  weakly locally finite \ $\Rightarrow$ \ $\B$  bi-locally finite.
\end{center}

The examples and the discussion given above show that there are bi-locally finite strong bimonoids which are not weakly locally finite. Up to now it has been  an open question whether each weakly locally finite strong bimonoid is even locally finite. Our first main result is:

\begin{theorem}\rm \label{thm:main-result-on-B}  There exists a right-distributive, weakly locally finite strong bimonoid $\B$  which is not locally finite.
\end{theorem}

We will prove a slightly stronger version of this result in Theorem \ref{thm:M-is-what-we-want}.
In our construction, we start with the free algebra of terms, with addition and multiplication as operations, over a set  $X$. By a natural congruence, the quotient of the term algebra becomes an additively locally finite and right-distributive strong bimonoid. We then construct a suitable further congruence which ensures that the corresponding quotient strong bimonoid  $\B$  is also multiplicatively locally finite; due to right-distributivity, it follows that  $\B$  is weakly locally finite.
The difficulty is to show that this finitely generated strong bimonoid is infinite; then it is clearly not locally finite. The latter is due to the condition ensuring multiplicative local finiteness and due to the lack of left-distributivity.

As mentioned, for all wta $\cA$  over a string ranked alphabet and a weakly locally finite strong bimonoid,
the image $\im(\initialsem{\cA})$  of its initial algebra semantics is finite. In view of Theorem~\ref{thm:main-result-on-B}, the question arises whether the corresponding statement holds for all wta over more arbitrary ranked alphabets and weakly locally finite strong bimonoids.
Our second main result shows that this statement fails drastically as soon as the ranked alphabet contains a binary symbol. In fact, there exist wta over such a ranked alphabet which have the universality property:

\begin{theorem}\rm \label{thm:main-result-on-Sigma} Let  $\Sigma$  be an arbitrary ranked alphabet
containing a binary symbol, and let  $\B$  be any finitely generated strong bimonoid.
Then there exists a weighted tree automaton $\cA$  over  $\Sigma$  and  $\B$  such that $\im(\initialsem{\cA}) = B$.
\end{theorem}

Note that the values of $\im(\initialsem{\cA})$  are given by iterated sums and products of the finitely many weights occurring in the wta $\cA$. By the above result, there are wta  $\cA$  which generate, in this way, \emph{all} values of the underlying strong bimonoid  $\B$  (provided it satisfies the necessary assumption of being finitely generated).
In particular, trivially, if $\B$  is infinite, the image  $\im(\initialsem{\cA})$  is also infinite.
By Theorem~\ref{thm:main-result-on-B}, this may happen also if $\B$  is weakly locally finite,
showing that weighted tree automata over ranked alphabets containing a binary symbol
have much larger generating power than weighted string automata.

The reason for the difference of this generative power with respect to the string case is that, for trees involving a binary symbol,
the calculation of the values of  $\initialsem{\cA}$  involves also \emph{products of sums of weights}
occurring in  $\cA$  or calculated along the evaluation on a tree, at proper branching points;
as we show, thereby infinitely many values and even all elements of  $\B$  can be produced.

Theorem~\ref{thm:main-result-on-Sigma} will be proved as Theorem \ref{lm:closure-of-finite-set-i-recognizable} . Our proof is constructive and proceeds roughly as follows.
By the assumption, the strong bimonoid  $\B$  is generated by a finite subset  $A \subseteq B$.
Given this finite set  $A$, we construct the wta  $\cA$  such that all its weights are from  $A \cup \{ \0, \1 \}$.
The elements generated by  $A \cup \{ \0, \1 \}$  can be represented by suitable trees over an auxiliary ranked alphabet with two binary symbols (modeling addition and multiplication, respectively). The wta  $\cA$  is constructed so that its transition functions reflect these operations; this can be done using only a single binary symbol from  $\Sigma$. Then, given any  $b \in B$, we obtain via its generation from  $A \cup \{ \0, \1 \}$  a suitable tree  $t \in \T_\Sigma$  for which, due to the construction of the transition functions, a calculation shows that $\initialsem{\cA}(t) = b$. This implies the result.

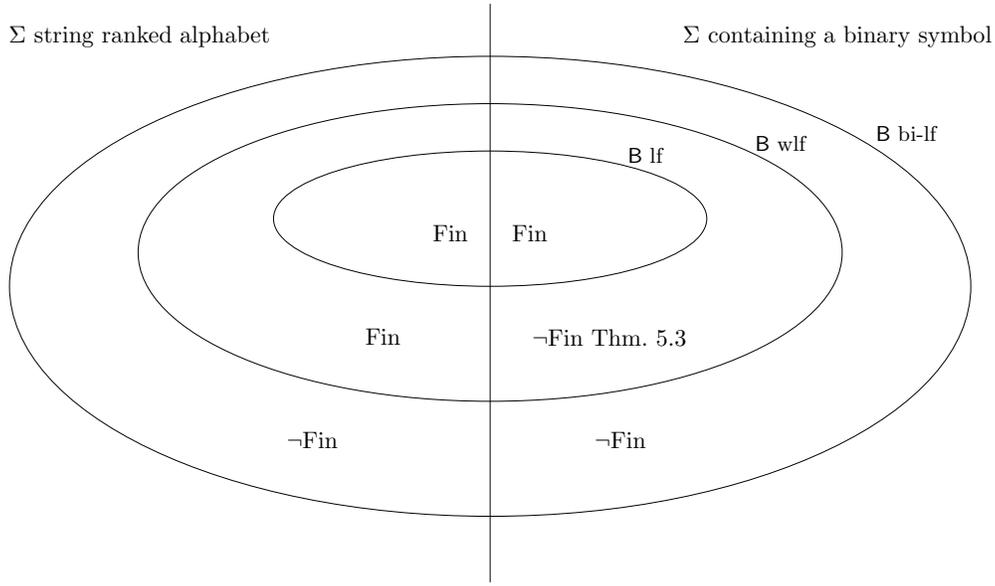
\begin{figure}
       \centering
         \def\bilocfinellipse{(0,0) ellipse (7.1cm and 3.4cm)} 
       \def\wlocfinellipse{(0,0.5) ellipse (5.2cm and 2.2cm)} 
       \def\locfinellipse{(0,1) ellipse (3.2cm and 1cm)} 
       \scalebox{0.9}{
       \begin{tikzpicture}
         \draw\bilocfinellipse node[label={[font=\small, label distance=5.8cm]20:$\sfB$\ bi-lf},
         label={[label distance=3.0cm,xshift=0.3cm]220:$\neg \mathrm{Fin}$},
         label={[label distance=3.0cm,xshift=-1.0cm]320:$\neg \mathrm{Fin}$ 
         }] {}; 
         %
         \draw\wlocfinellipse node[label={[font=\small, label distance=3.9cm]20:$\sfB\ \mathrm{wlf}$},
         label={[label distance=1.4cm]220:$\mathrm{Fin}$},
         label={[label distance=1.4cm, xshift=-0.7cm]320:$\neg \mathrm{Fin}$ Thm.~\ref{cor:second-main}}]{}; 
         %
         \draw\locfinellipse node[label={[font=\small, label distance=1.9cm]20:$\sfB\ \mathrm{lf}$},
         label={[label distance=0.1cm, yshift=0.2cm]220:$\mathrm{Fin}$},
         label={[label distance=0.1cm, yshift=0.2cm]320:$\mathrm{Fin}$}] {}; 

\node at (0,4.3) (upper-node) {};
\node at (0,-4.5) (lower-node) {};
\draw (upper-node) -- (lower-node);
\node[below left=0.2cm and 3cm of upper-node]{$\Sigma$ string ranked alphabet};
\node[below right=0.2cm and 2.6cm of upper-node]{$\Sigma$ containing a binary symbol};
\end{tikzpicture}
}
       \caption{\label{fig:overview-finiteness} An overview  for finite image property with respect to initial algebra semantics. In the figure lf = locally finite, wlf = weakly locally finite, and bi-lf = bi-locally finite. Moreover, Fin  means that, for each wta  $\cA$  over such an alphabet and such a strong bimonoid  $\B$,
$\im(\initialsem{\cA})$  is finite. 
}
       \end{figure}
       
 To the best of our knowledge, Theorem~\ref{thm:main-result-on-Sigma} is new also for the case of semirings.
As mentioned before, for semirings, the run semantics and the initial algebra semantics coincide.
Hence, as an immediate consequence of this and Theorem~\ref{thm:main-result-on-Sigma}, we obtain that for each ranked alphabet
$\Sigma$ containing a binary symbol and for each finitely generated semiring  $\sfB=(B,\oplus,\otimes,\0,\1)$, there exists a wta $\cA$
over  $\Sigma$  and  $\sfB$  such that $\im(\runsem{\cA}) = B$.
Note that we obtained this result as a consequence of the construction for the initial algebra semantics.
It would be interesting to see whether a direct proof for the run semantics is possible.

The diagram in Figure \ref{fig:overview-finiteness} summarizes the results on the finite image property of wta with initial algebra semantics. It assumes an arbitrary
string ranked alphabet  $\Sigma$,  respectively, ranked alphabet $\Sigma$  containing a binary symbol to be given.
The question is whether then, for each strong bimonoid  $\B$  which is locally finite, weakly locally finite,
or bi-locally finite and for each wta  $\cA$  over  $\Sigma$ and  $\B$, the image of the initial algebra semantics
$\im(\initialsem{\cA})$  is finite (this is expressed by 'Fin'). The part missing up to now concerned the case that
$\Sigma$  contains a binary symbol and  $\B$  is weakly locally finite. This is filled by Theorem \ref{cor:second-main}, 
a consequence of Theorems \ref{thm:main-result-on-B}  and \ref{thm:main-result-on-Sigma},
thereby completing the picture.

Open research questions are presented and discussed in Section \ref{sect:further-research}.

\section{Preliminaries}
\label{sec:preliminaries}

\paragraph{General.} We denote by $\mathbb{N}$  the set of nonnegative integers and $\mathbb{N}_+=\mathbb{N}\setminus\{0\}$. For every $k,n \in \mathbb{N}$, we denote by $[k,n]$ the set $\{i\in \mathbb{N} \mid k \le i \le n\}$ and we abbreviate $[1,n]$ by $[n]$. Hence $[0] = \emptyset$.

Let $A$ be a set.  We denote by $A^*$ the set of all finite sequences over $A$, and by $\varepsilon$ the empty sequence. 

Let $\rho \subseteq A\times A$ be a binary relation on $A$. As usual, $\rho^*$ denotes the reflexive and transitive closure of $\rho$. (It will always be clear from the context whether $\rho^*$ denotes the reflexive and  transitive closure of the relation $\rho$ or the set of all finite sequences over the set $\rho$.) The relation $\rho$ is
 an \emph{equivalence relation on $A$} if it is reflexive, symmetric, and transitive.
If this is the case, then for each $a\in A$, the \emph{equivalence class of $a$ (modulo $\rho$)}, denoted by $[a]_\rho$, is the set $\{b\in A \mid a\rho b\}$. For each $B\subseteq A$, we put $B/_\rho =\{[a]_\rho \mid a\in B\}$.

\paragraph{Ranked alphabets and terms.}    A {\em ranked alphabet} is a pair $(\Sigma,\rk)$, where $\Sigma$ is a non-empty and finite set and $\rk: \Sigma \rightarrow \mathbb{N}$ is a mapping, called \emph{rank  mapping},  such that $\rk^{-1}(0)\not= \emptyset$. For each $k\in \mathbb{N}$, we denote the set $\rk^{-1}(k)$ by $\Sigma^{(k)}$. Sometimes we write $\sigma^{(k)}$ to indicate that $\sigma \in \Sigma^{(k)}$. Moreover, we abbreviate $(\Sigma,\rk)$ by $\Sigma$ and assume that the rank mapping is known or irrelevant.

A ranked alphabet $\Sigma$ is \emph{monadic} if $\Sigma=\Sigma^{(0)} \cup \Sigma^{(1)}$. A monadic ranked alphabet is 
a \emph{string ranked alphabet} if $|\Sigma^{(0)}|=1$ and $\Sigma^{(1)}\not= \emptyset$.

Let $\Sigma$ be a ranked alphabet and $X$ be a set disjoint with $\Sigma$.
The set of \emph{$\Sigma$-terms (or $\Sigma$-trees) over $X$, } denoted by $\T_\Sigma(X)$, is the smallest set $T$ such that (i) $\Sigma^{(0)} \cup X \subseteq T$ and (ii) for every $k \in \mathbb{N}_+$, $\sigma \in \Sigma^{(k)}$, and $t_1,\ldots, t_k \in T$, we have $\sigma(t_1,\ldots,t_k) \in T$. We abbreviate $\T_\Sigma(\emptyset)$ by $\T_\Sigma$.

We note that the name $\Sigma$-term (or simply term) is used, among others,  in algebra and in the theory of term rewriting. At the same time, 
we write and say $\Sigma$-tree (or simply tree) in the theory of tree automata and tree languages. Since in the present paper we deal with all these mentioned 
areas, we will write term in Section  \ref{sect:wlc-strong-bimonoids} and tree in Sections \ref{sect:wta-def} and \ref{sect:wta-section}.

\paragraph{Universal algebra.} 
  In the following, we recall some concepts and results from universal algebra which can be found e.g. in \cite{bursan81,wec92}, and \cite{baanip98}.
In universal algebra, a ranked alphabet is called a signature. Each letter of the ranked alphabet is an operation symbol of the same arity. Nullary letters (leaves) are nullary operations (or constants). 

Let $\Sigma$ be a ranked alphabet (signature). An \emph{algebra $\sf A$ of type $\Sigma$} ($\Sigma$-algebra) is a pair $\sfA=(A,\theta)$ where $A$ is a nonempty set and $\theta$ is a mapping from $\Sigma$ to the family of finitary operations on $A$ such that for every $k\in \mathbb{N}$ and  $\sigma \in \Sigma^{(k)}$, the arity of the operation $\theta(\sigma)$ is $k$. In particular, for $k=0$, a nullary operation is a mapping of type $A^0 \to A$ where $A^0=\{()\}$. 

Let $X$ be a set. The \emph{$\Sigma$-term algebra over $X$}, denoted by $\sfT_\Sigma(X)$,  is the $\Sigma$-algebra
\(\sfT_\Sigma(X) = (\T_\Sigma(X),\theta_\Sigma)\) where, for every $k \in \mathbb{N}$, $\sigma \in \Sigma^{(k)}$, and $t_1,\ldots, t_k \in \T_\Sigma(X)$, we let $\theta_\Sigma(\sigma)(t_1,\ldots,t_k) = \sigma(t_1,\ldots,t_k)$. The \emph{$\Sigma$-term algebra}, denoted by $\sfT_\Sigma$, is the $\Sigma$-term algebra over $\emptyset$, i.e., $\sfT_\Sigma = \sfT_\Sigma(\emptyset)$.

Let $\sfA=(A,\theta)$ be a $\Sigma$-algebra and $A' \subseteq A$. We say that \emph{$A'$ is closed under $\theta(\Sigma)$} if, for every $k \in \mathbb{N}$,  $\sigma \in \Sigma^{(k)}$, and $a_1,\ldots,a_k \in A'$, we have $\theta(\sigma)(a_1,\ldots,a_k) \in A'$. We denote by $\langle A' \rangle_{\theta(\Sigma)}$ the smallest subset of $A$ which contains $A'$ and is closed under $\theta(\Sigma)$. The \emph{subalgebra of $\sfA$ generated by $A'$} is the $\Sigma$-algebra $( \langle A' \rangle_{\theta(\Sigma)},\theta')$ where $\theta'$ is obtained from $\theta$ by restricting each operation $\theta(\sigma)$ to $\langle A' \rangle_{\theta(\Sigma)}$.
If $A=\langle A' \rangle_{\theta(\Sigma)}$, then $\sfA$ is \emph{generated by $A'$}.

The $\Sigma$-algebra $\sfA=(A,\theta)$ is \emph{locally finite} if, for each finite subset $A' \subseteq A$, the set $\langle A' \rangle_{\theta(\Sigma)}$ is finite.

Let $\sfA_1=(A_1,\theta_1)$ and $\sfA_2=(A_2,\theta_2)$ be $\Sigma$-algebras. A \emph{$\Sigma$-algebra homomorphism (from $\sfA_1$ to $\sfA_2$)}  is a mapping $h: A_1 \to A_2$ such that, for every $k \in \mathbb{N}$, $\sigma \in \Sigma^{(k)}$, and $a_1,\ldots,a_k \in A_1$, we have $h(\theta_1(\sigma)(a_1,\ldots,a_k)) = \theta_2(\sigma)(h(a_1),\ldots,h(a_k))$.

  \label{page:initial-algebra}
Let $\mathcal{K}$ be an arbitrary class of $\Sigma$-algebras. Moreover, let $\A=(A,\theta)$ be a $\Sigma$-algebra in $\mathcal{K}$ and $X \subseteq A$ such that $\A$ is generated by $X$. The algebra~$\A$ is called \emph{free in~$\mathcal{K}$ with generating set $X$} if, for every $\Sigma$-algebra $\A'=(A',\theta')$ in $\mathcal{K}$ and mapping $f\colon X \rightarrow A'$, there exists a unique extension of $f$ to a $\Sigma$-algebra homomorphism $h\colon A \rightarrow A'$ from $\A$ to $\A'$. The algebra $\sfT_\Sigma(X)$ is free in the class of all $\Sigma$-algebras with generating set $X$  \cite[Thm.~II.10.8]{bursan81}, \cite[p.~18, Thm.~4]{wec92}.
If $\A$ is free in~$\mathcal{K}$ with generating set $X=\emptyset$, then for each~$\A'=(A',\theta')$ in $\mathcal{K}$, there exists exactly one $\Sigma$-algebra homomorphism $h\colon A \rightarrow A'$  from $\A$ to $\A'$. In this case $\A$ is called \emph{initial in~$\mathcal{K}$} \cite[p.~164, Def.~4]{wec92}.
 The $\Sigma$-term algebra $\sfT_\Sigma$ is initial in the class of all $\Sigma$-algebras.

In the rest of this section, $\sfA$ denotes an arbitrary $\Sigma$-algebra $(A,\theta)$.

A \emph{congruence (relation) on $\sfA$} is an equivalence relation $\rho\subseteq A\times A$ which satisfies the following condition: for every $k\in \mathbb{N}$, $\sigma \in \Sigma^{(k)}$, $a_1,b_1,\ldots,a_k,b_k \in A$, if $a_i \,\rho\, b_i$ for each $i\in[k]$, then
\[\theta(\sigma)(a_1,\ldots,a_k ) \, \rho \, \theta(\sigma)(b_1,\ldots,b_k ) .\]

Let  $\rho$ be a congruence on $\sfA$. The \emph{quotient algebra of $\sfA$ by $\rho$} is the $\Sigma$-algebra $\sfA/\!_\rho = (A/\!_\rho, \theta/\!_\rho)$ where, for every $k \in \mathbb{N}$,  $\sigma \in \Sigma^{(k)}$, and $[a_1]_{\rho},\ldots,[a_k]_{\rho} \in A/\!_\rho$, we have $\theta/\!_\rho(\sigma)([a_1]_{\rho},\ldots,[a_k]_{\rho}) =  [\theta(\sigma)(a_1,\ldots, a_k)]_{\rho}$.

Next we wish to consider $\Sigma$-identities and the congruence on $\sfA$ induced by a set of such identities. For this, we introduce the necessary concepts. 

Let $Z=\{z_1,z_2,\ldots\}$ be a set of variables. For each $n\in \mathbb{N}$, we put $Z_n=\{z_1,\ldots,z_n\}$.  

Let $t\in \T_\Sigma(A\cup Z_n)$ for some $n\in \mathbb{N}$. The mapping $t^\sfA: A^n \to A$ is defined by structural induction in a standard way. We note that such a mapping is called 
algebraic function in \cite[p.~22]{gecste84} and term function in \cite[Def.~II.10.2]{bursan81} for the special case that $t\in \T_\Sigma(Z_n)$.

Each element $t\in \T_\Sigma(A\cup Z_1)$ in which $z_1$ occurs exactly once is called a {\em $\Sigma A$-context}. The set of all $\Sigma A$-contexts is denoted by $\C_{\Sigma,A}$ (cf. unary algebraic functions in  \cite[Def.~1.3.13]{gecste84}). 

An \emph{assignment} is a mapping $\varphi: Z \to A$. Each such mapping $\varphi$ extends uniquely to a $\Sigma$-algebra homomorphism
$\varphi'$ from  $\sfT_\Sigma(A\cup Z)$ to $\sfA$ satisfying that $\varphi'(a)=a$ for each $a\in A$. In the sequel, we drop the prime from  $\varphi'$. For an arbitrary $t \in \T_\Sigma(A\cup Z)$, we call $\varphi(t)$ the {\em evaluation of $t$ in $ \sfA$ at $\varphi$}. 

We note that for each $n\in \mathbb{N}$, $t\in \T_\Sigma(A\cup Z_n)$, and assignment $\varphi$ with $\varphi(z_i)=a_i$ for $i\in [n]$, we have $\varphi(t)=t^{\sfA}(a_1,\ldots,a_n)$.

A \emph{$\Sigma$-identity over $Z$} (or: identity) is a pair $(\ell, r)$ where $\ell,r \in \T_\Sigma(Z)$.  The $\Sigma$-algebra $\sfA$ \emph{satisfies the identity $(\ell, r)$} if, for every assignment $\varphi: Z \to A$, we have
$\varphi(\ell) = \varphi(r)$.

\begin{lemma}\rm \label{identitiestofactoralgebras}\cite[Th.~II.6.10 and Lm.~II.11.3]{bursan81}
If $\sfA$ satisfies an identity $(\ell, r)$ and $\rho$ is a congruence on $\sfA$, then $\sfA/\!_\rho$ also satisfies the identity $(\ell, r)$. 
\end{lemma}

Let $E$ be a set of identities. The \emph{congruence (relation)  on $\sfA$ induced by $E$}, denoted by $\approx_E$, is the smallest congruence on $\sfA$ which contains the set
\begin{equation}\label{eq:identity-in-term-algebra}
E(\sfA)=\{(\varphi(\ell), \varphi(r))\mid (\ell, r)\in E, \varphi: Z \to A\}.
\end{equation}

The following lemma is well-known and can be proven similarly to \cite[p.~176, Lm.~24]{wec92}.

\begin{lemma}\label{lm:free-algebra-quotient} \rm Let  $E$ be a set of $\Sigma$-identities. Then $\sfA/\!_{\approx_E}$ satisfies all identities in $E$.
\end{lemma}

Next we extend the well-known syntactic characterization of the congruence on $\sfT_\Sigma(Z)$ induced by a set $E \subseteq \T_\Sigma(Z) \times \T_\Sigma(Z)$ of $\Sigma$-identities, 
cf. \cite[Thm.~ 3.1.12]{baanip98} and \cite[Thm.~II.14.17, II.14.19]{bursan81},
to a characterization of the congruence on $\sfA$ induced by  $E$. In fact, this is closely related to a general description of a congruence generated by a binary relation on  $\A$,
cf. \cite[Sect.~2.1.2]{wec92}.

Let $E$ be a set of $\Sigma$-identities. The \emph{reduction relation induced by $E$ on $A$}, denoted  by  $\Rightarrow_E$, is the binary relation on $A$ defined as follows:
for every $a,b \in A$, we let $a \Rightarrow_E b$ if there exist a $\Sigma A$-context $c\in \C_{\Sigma,A}$, an identity $(\ell, r)$ in $E$,  and an assignment $\varphi: Z \to A$ such that
$a= c^\sfA(\varphi(\ell))$ and $b = c^\sfA(\varphi(r))$.
In this  case  we say that $b$ is obtained from $a$  in a reduction step (using the identity $(\ell, r)$).

For an identity $e=(\ell,r)$ we define $e^{-1}=(r,\ell)$ and we let $E^{-1}=\{ e^{-1} \mid e\in E\}$.
Moreover, we abbreviate $\Rightarrow_{E\cup E^{-1}}$ by $\Leftrightarrow_E$.

The subsequent characterization says that, for any two elements $a,b\in A$,
we have $a \approx_E b$ if and only if, there is a finite sequence of elements $a=a_0,a_1,\ldots,a_n=b$ of $A$ for some $n\in \mathbb{N}$ such that for each $i\in [n]$, the element $a_i$ can be obtained from $a_{i-1}$ in a reduction step using an identity in $E$  or the inverse of an identity. As it will be crucial for us (cf. the proof of Theorem \ref{thm:M-is-what-we-want}), we include a proof for the convenience of the reader, in our present notation. 

\begin{lemma}\rm \label{lm:approx-characterization}\  \cite[p.~98, Thm.~6]{wec92} Let  $E$ be a set of $\Sigma$-identities and $\approx_E$ the congruence on $\sfA$ induced by $E$. Then $\approx_E \, =\, \Leftrightarrow^*_E$.
\end{lemma}
\begin{proof} As a first step, we show that $\Leftrightarrow^*_E$ is a congruence on $\sfA$.

For this, let  $a, b \in A$  with  $a \Leftrightarrow^*_E b$, $k\ge 1$,  $\sigma \in \Sigma^{(k)}$,  $i\in [k]$ and $a_1,\ldots, a_{i-1},a_{i+1},\ldots,a_k \in A$.
We show that 
\[\theta(\sigma)(a_1,\ldots,a_{i-1},a,a_{i+1},\ldots,a_k) \Leftrightarrow^*_E  \theta(\sigma)(a_1,\ldots,a_{i-1},b,a_{i+1},\ldots,a_k).\]
For the sake of simplicity we assume that $i=1$, the proof for an arbitrary $i\in [k]$ is similar.

First assume that  $a \Rightarrow_E b$. 
Then there exist a $\Sigma A$-context $c\in \C_{\Sigma,A}$, an identity $(\ell, r)$ in $E$,  and an assignment $\varphi: Z \to A$ such that
$a= c^\sfA(\varphi(\ell))$ and $b = c^\sfA(\varphi(r))$. Then for the $\Sigma A$-context $c'=\sigma(c,a_2,\ldots,a_k)$ we have $ (c')^\sfA(\varphi(\ell)) \Rightarrow_E (c')^\sfA(\varphi(r))$.
Moreover,
\[ (c')^\sfA(\varphi(\ell)) = \theta(\sigma)(c^\sfA(\varphi(\ell)),a_2,\ldots, a_k)= \theta(\sigma)(a,a_2,\ldots,a_k)  \text{ and }\]
\[ (c')^\sfA(\varphi(r)) = \theta(\sigma)(c^\sfA(\varphi(r)),a_2,\ldots, a_k)= \theta(\sigma)(b,a_2,\ldots,a_k);\]
this proves that $\theta(\sigma)(a,a_2,\ldots,a_k) \Rightarrow_E \theta(\sigma)(b,a_2,\ldots,a_k)$. 

Using the above, by symmetry, we obtain that $a \Leftrightarrow_E b$ implies
$\theta(\sigma)(a,a_2,\ldots,a_k) \Leftrightarrow_E \theta(\sigma)(b,a_2,\ldots,a_k)$. Lastly, by an easy induction, we can show that, for every $n\in\mathbb{N}$, $a \Leftrightarrow^n_E b$ implies
$\theta(\sigma)(a,a_2,\ldots,a_k) \Leftrightarrow^n_E \theta(\sigma)(b,a_2,\ldots,a_k)$. Hence  $\Leftrightarrow^*_E$  is a congruence.

Now, since $\Leftrightarrow_E\,\subseteq \,\Leftrightarrow^*_E$ and $\Leftrightarrow_E$ contains the set $E(\sfA)$ by definition, we obtain $\approx_E \, \subseteq \, \Leftrightarrow^*_E$.

We claim that also  $\Leftrightarrow^*_E \subseteq \approx_E$. Since  $\approx_E$  is an equivalence relation, it suffices to show that  $\Rightarrow_E \subseteq \approx_E$.

Let $a,b \in A$ such that $a \Rightarrow_E b$. By the definition of $\Rightarrow_E$,
there are a context $c\in \C_{\Sigma,A}$, an identity $(\ell,r)\in E$, and an assignment $\varphi:Z \to A$  such that
$a=c^\sfA(\varphi(\ell))$  and  $b=c^\sfA(\varphi(r))$.
By \eqref{eq:identity-in-term-algebra}, 
we have $(\varphi(\ell), \varphi(r)) \in E(\sfA)$, hence
$ \varphi(\ell) \approx_E \varphi(r)$. Since $\approx_E$ is a congruence, we have 
$c^\sfA(\varphi(\ell))\approx_E c^\sfA(\varphi(r))$, 
i.e. $a\approx_E b$. 
\end{proof}

\paragraph{Strong bimonoids.}

\sloppy A \emph{strong bimonoid} \cite{drostuvog10,cirdroignvog10,rad10,drovog10,drovog12} is an algebra $\B=(B,\oplus,\otimes,\0,\1)$ such that $(B,\oplus,\0)$ is a commutative monoid, $(B,\otimes,\1)$ is a monoid, and $\0$ is annihilating with respect to $\otimes$, i.e., for each $b \in B$ we have $b\otimes \0 = \0 \otimes b = \0$. The operations $\oplus$ and $\otimes$ are called addition and multiplication, respectively. For examples of strong bimonoids we refer to \cite{drostuvog10,cirdroignvog10} (also cf. \cite[Ex.~2.7.10]{fulvog22}).

Let $\B=(B,\oplus,\otimes,\0,\1)$ be a strong bimonoid. It is
\begin{compactitem}
\item \emph{idempotent} if, for each $b \in B$, we have $b \oplus b = b$,
\item  \emph{almost idempotent} if, for each $b \in B$, we have $b \oplus b \oplus b = b \oplus b$,
  \item \emph{commutative} if $\otimes$ is commutative,
\item \emph{left-distributive} if, for every $a,b,c \in B$, we have $a \otimes (b \oplus c) = (a \otimes b) \oplus (a \otimes c)$,
\item \emph{right-distributive} if, for every $a,b,c \in B$, we have $(a \oplus b) \otimes c = (a \otimes c) \oplus (b \otimes c)$.
  \end{compactitem}
A \emph{semiring} \cite{hebwei93,gol99} is a distributive strong bimonoid, i.e., a strong bimonoid which is left-distributive and right-distributive.

Let $\B=(B,\oplus,\otimes,\0,\1)$ be a strong bimonoid and $A\subseteq B$. The \emph{weak closure of $A$ (with respect to $\B$)}, denoted by $\wcl_\B(A)$,  is the smallest subset $C\subseteq B$ such that $A\cup\{\0,\1\}\subseteq C$ and, for every  $b,b'\in C$ and $a \in A$, we have $(b\oplus b') \in C$ and $(b\otimes a) \in C$. If $\B$ is clear from the context, then we drop $\B$ from $\wcl_\B(A)$ and simply write $\wcl(A)$. 

We call the strong bimonoid $\B$
\begin{compactitem}
  \item \emph{additively locally finite} if  $(B,\oplus,\0)$ is locally finite,
  \item \emph{multiplicatively locally finite} if  $(B,\otimes,\1)$ is locally finite,
  \item \emph{bi-locally finite} if it is additively and multiplicatively locally finite, and
  \item \emph{weakly locally finite} if, for each finite subset $A \subseteq B$, the weak closure of $A$ is finite.
  \end{compactitem}
  
  Next we present some properties of strong bimonoids which we will use in the paper.

\begin{observation}\label{obs:biloc-fin+right-disrt-loc-fin} \rm   Let $\B=(B,\oplus,\otimes,\0,\1)$ be a strong bimonoid.  
  \begin{compactitem}
  \item[(a)] If $\B$ is locally finite, then it is weakly locally finite. If $\B$ is weakly locally finite, then it is bi-locally finite.
 \item[(b)] \cite[Rem.~17]{drostuvog10} If $\B$ is right-distributive, then it is bi-locally finite if and only if it is weakly locally finite.
 \item[(c)] If $\B$ is almost idempotent, then it is additively locally finite.
      \end{compactitem}
\end{observation}
\begin{proof} We give a short proof only for (c). Let  $A \subseteq B$  be a finite subset.
Then  $\langle A\rangle_{\{\oplus,\0\}}$  consists of all finite sums of elements from  $A$
in which each summand occurs at most twice. Hence  $\langle A\rangle_{\{\oplus,\0\}}$  is finite.
\end{proof}

\begin{example}\rm \cite[Ex.~2.1(2)]{drovog12} (cf. \cite[Ex.~2.6.10(2)]{fulvog22}) \label{ex:Trunc} For each $\lambda \in \mathbb{R}$ with $0 < \lambda< \frac{1}{2}$,  let $\Trunc_\lambda= (B,\oplus,\odot,0,1)$ be the algebra, where
  \begin{compactitem}
    \item $B =\{0\} \cup \{b \in \mathbb{R}\mid \lambda \le b \le 1\}$,
    \item $a \oplus b = \min(a + b, 1)$,  and 
    \item $a\odot b  = a \cdot b$ if $a \cdot b \ge \lambda$, and $0$ otherwise,
    \end{compactitem}
    and where $+$ and $\cdot$ are the usual addition and multiplication of real numbers, respectively.

   Obviously, $\Trunc_\lambda$ is a commutative and bi-locally finite strong bimonoid. It is not a semiring because $\odot$ is not right-distributive. For instance, for $a=b=0.9$, and $c=\lambda$, we have $(a \oplus b) \odot c=\lambda$, while $(a \odot c) \oplus (b \odot c)=0$ because $a \odot c = b \odot c = 0$.
   
   We show that $\Trunc_{\frac{1}{4}}$ is not weakly locally finite. For this, we define the family $(b_n \mid n \in \mathbb{N})$ such that $b_0=\frac{1}{2}$ and, for each $n \in \mathbb{N}_+$, we let $b_n = b_{n-1} \cdot \frac{1}{2}$ if $n$ is odd, and $b_n = b_{n-1} + \frac{1}{2}$ if $n$ is even. Thus, e.g., $b_0=\frac{1}{2}$, $b_1=\frac{1}{4}$, $b_2=\frac{3}{4}$, $b_3=\frac{3}{8}$, and $b_4=\frac{7}{8}$.
Then $\{b_n \mid n \in \mathbb{N}\} \subseteq \mathrm{wcl}(\{\frac{1}{2}\})$. Clearly,
 \begin{equation}\label{equ:bn-even-odd}
        \text{for each $n \in \mathbb{N}$, if $n$ is even, then $1/2 \le b_n \le 1$, and if $n$ is odd, then $1/4 \le b_n < 1/2$.}
        \end{equation}
Hence $b_i\not= b_j$ for every $i,j \in \mathbb{N}$ with $i \not= j$, and thus  $\{b_n \mid n \in \mathbb{N}\}$ is not finite. This means that $\mathrm{wcl}(\{\frac{1}{2}\})$ is not finite and $\Trunc_{\frac{1}{4}}$ is not weakly locally finite.    \hfill $\Box$
\end{example}

\begin{example}\rm \label{ex:Stb} \cite[Ex.~25]{drostuvog10}\footnote{$\mathsf{Stb}$ refers to one of the authors} (also cf. \cite[Ex.~2.6.10(9)]{fulvog22}) In the strong bimonoid $\mathsf{Stb}=(\mathbb{N},\oplus,\odot,0,1)$, let the two commutative operations $\oplus$ and $\odot$ on
        $\mathbb N$ satisfy the following requirements.  If $a,b\in
        \mathbb N\setminus\{0\}$ with $a\leq b$, we have (with $+$
        being the usual addition on $\mathbb{N}$) 
	\begin{align*}
		a\oplus b &=
		\begin{cases}
			b & \text{if $b$ is even}\\
			b+1 & \text{if $b$ is odd.} 
		\end{cases}
	\intertext{If $a,b\in\mathbb N\setminus\{0,1\}$ with $a\leq b$, then}
		a\odot b &=
		\begin{cases}
			b+1 &\text{if $b$ is even}\\
			b & \text{if $b$ is odd.} 
		\end{cases}
	\end{align*}
Clearly, $\mathsf{Stb}$ is not a semiring because, e.g., $2 \odot (2 \oplus 3) = 2 \odot 4 = 5$ and $(2\odot 2) \oplus (2\odot 3) = 3 \oplus 3 = 4$. Also, it is easy to see that $\mathsf{Stb}$ is bi-locally finite.       
        However, if we apply $\oplus$ and $\odot$ alternatingly, then the result increases arbitrarily. 
        For instance, let $(b_i \mid n \in \mathbb{N})$ be the family defined by $b_0 = 2$ and, for each $n \in \mathbb{N}$, by
        \[
          b_{n+1} = \begin{cases}
            b_n \oplus 2 & \text{ if $n$ is odd}\\
            b_n \odot 2 & \text{ otherwise}\enspace.
            \end{cases}
          \]
          Then, e.g., $b_0 = 2$, $b_1 = 3$, $b_2 = 4$, $b_3=5$, .... Hence $\mathsf{Stb}$ is not weakly locally finite.  \hfill $\Box$
  \end{example}


\section{Weakly locally finite strong bimonoids}\label{sect:wlc-strong-bimonoids}

The main goal of this section is to show that there exists a right-distributive strong bimonoid which is weakly locally finite but not locally finite, cf. Theorem~\ref{thm:M-is-what-we-want}.

For the proof, for each nonempty set $X$, we define an almost idempotent right-distributive strong bimonoid $\sfM(X)$ and prove that it is weakly locally finite  and not locally finite. We define $\M(X)$ in two steps. In the first step, we define an algebra  $\sfST_\Sigma(X)$ which looks very similar to the free  $\Sigma$-algebra  $\sfT_\Sigma(X)$, but incorporates
the usual laws for  $0$  and $1$. Then the quotient
$\sfS(X) = \sfST_\Sigma(X)/_{\approx_E}$  by the congruence induced by a set  $E$ of natural identities is a strong bimonoid which is right-distributive and almost idempotent (see Lemma
  \ref{prop:SB-strong-bimonoid}). 
  In the second step, we factorize $\sfS(X)$ by the congruence relation $\sim_\mathrm{la}$ which identifies multiplicatively ``large'' elements of $\sfS(X)$ (cf. Definition~\ref{def:large-polynomial}) in order to obtain the multiplicatively locally finite algebra $\M(X)$. Since $\M(X)$ is almost idempotent and hence additively locally finite, we obtain from Observation~\ref{obs:biloc-fin+right-disrt-loc-fin}(b) that $\M(X)$ is weakly locally finite. In Theorem~\ref{thm:M-is-what-we-want} we prove that $\M(X)$ is not locally finite.
  This uses Lemma \ref{lm:approx-characterization} and exploits
our choice of the identities of  $E$, namely, that particular
constructed terms permit only one reduction rule (or its inverse) from  $E$.

  \begin{quote}\emph{In the rest of this section, we let $\Sigma = \{+^{(2)}, \times^{(2)}, 0^{(0)}, 1^{(0)}\}$ and let $X$ be a nonempty set.}
    \end{quote}
  
\underline{Step 1:} We write elements of $\T_\Sigma(X)$ in infix form, e.g., we write $(1+1)\times x$ for $\times(+(1,1),x)$ where $x\in X$.

We call a term  $t  \in \T_\Sigma(X)$  \emph{simple}, if  
\begin{compactitem}
\item $t = 0$  or
\item $t\ne 0$  and it contains neither   $0$, nor a subterm of the form  $1 \times t$,  nor a subterm of the form $t \times 1$.
\end{compactitem}Let $\rmST(X)$ denote the set of all simple terms in $\T_\Sigma(X)$. Note that $1$ is simple, hence
e.g., $1+t\in\rmST(X) $ for each $t\in \rmST(X)$. 

We define the algebra $\sfST_\Sigma(X)=(\rmST(X),+_\mathsf{ST},\times_\mathsf{ST},0,1)$ as follows:
\begin{compactitem}
\item for each $t\in \rmST(X)$, let $t+_\mathsf{ST} 0= 0 +_\mathsf{ST} t=t$ and   $t \times_\mathsf{ST} 0= 0 \times_\mathsf{ST} t=0$,
\item for each $t\in \rmST(X)$, let $t\times_\mathsf{ST} 1= 1 \times_\mathsf{ST} t=t$,
\item for every $s,t\in \rmST(X)\setminus\{0,1\}$, let $s +_\mathsf{ST} t = s+t$ and $s \times_\mathsf{ST} t = s\times t$.
\end{compactitem}

Next let $E$ be the set of the following five identities:
\begin{center}
\begin{tabular}{ll}
$e_1: \big(z_1 + (z_2 + z_3) \ , \ (z_1 + z_2) + z_3\big)$ \hspace{8mm} & $e_4: \big(z_1 + z_1 \ , \ z_1 + (z_1 + z_1) \big)$ \\[2mm]
$e_2: \big(z_1 + z_2 \ , \ z_2 + z_1\big)$ & $e_5:  \big((z_1 + z_2)\times z_3 \ , \ (z_1 \times z_3) + (z_2 \times z_3) \big)$ \\[2mm]
$e_3: \big(z_1 \times (z_2 \times z_3)  \ , \  (z_1 \times z_2) \times z_3\big)$  
\end{tabular}
\end{center}
We note that $E \subseteq \T_\Sigma(Z_3) \times \T_\Sigma(Z_3)$.

Then we consider the quotient algebra 
\[\sfST_\Sigma(X)/_{\approx_E}=(\rmST(X)/_{\approx_E},+_\mathsf{ST}/_{\approx_E},\times_\mathsf{ST}/_{\approx_E},[0]_{\approx_E},[1]_{\approx_E}).\]
Lastly, we abbreviate the latter notation by $\sfS(X)=(\rmS(X),+_\sfS,\times_\sfS,0_\sfS,1_\sfS)$, and,  for each $s\in \rmST(X)$, we abbreviate $[s]_{\approx_{E}}$ by $[s]_{E}$.

\begin{lemma}\rm \label{prop:SB-strong-bimonoid} The algebra $\sfS(X)$ is an almost idempotent right-distributive strong bimonoid.
\end{lemma}
\begin{proof} By the definition of $+_\mathsf{ST}$ and $\times_\mathsf{ST}$,  the terms 0 and 1 are the additive unit element and multiplicative unit element in $\sfST_\Sigma(X)$, respectively. Moreover, 0 is annihilating with respect to $\times_\mathsf{ST}$. Each of these properties can be described by a corresponding identity satisfied by  $\sfST_\Sigma(X)$. By Lemma \ref{identitiestofactoralgebras}, $\sfS(X)$ also satisfies these identities. Hence, $0_\sfS$ and $1_\sfS$ are the additive unit element and multiplicative unit element in $\sfS(X)$, respectively, and $0_\sfS$ is annihilating with respect to  $\times_\sfS$.

Moreover, by Lemma \ref{lm:free-algebra-quotient} (with $\sfA=\sfST_\Sigma(X)$), the algebra $\sfS(X)$ satisfies the identities  $e_1 - e_5$. Identities $e_1 - e_3$, together with the mentioned properties of $0_\sfS$ and $1_\sfS$, ensure that  $\sfS(X)$  is a strong bimonoid, then identities $e_4$ and $e_5$ ensure that it is almost idempotent and right-distributive, respectively.
\end{proof}

\underline{Step 2:} Clearly,  $\sfS(X)$  is not multiplicatively locally finite, since for each  $x \in X$,
the multiplicative submonoid  $\langle\{[x]_E\}\rangle_{\{\times_{\sfS},1_\sfS\}}$  generated by  $\{[x]_E\}$  is infinite.
In order to construct a factor algebra that is
multiplicatively locally finite we define the concept  of a large element in  $\rmS(X)$. Intuitively, all products $ p \times_{\sfS} (q \times_{\sfS} r)$ of elements $p, q, r \in \rmS(X) \setminus \{ 0_\sfS, 1_\sfS \}$ are large, and any element obtained from a large element by adding or multiplying it with any further element should also be large.
The formal definition is the following.

\begin{definition}\label{def:large-polynomial} \rm An element $p \in \rmS(X) \setminus \{ 0_\sfS, 1_\sfS \}$  is \emph{large},
if  there is a term $s\in \rmST(X)$ such that:
\begin{compactitem}
\item[(a)]  $p = [s]_{E}$  and 
\item[(b)]  $s$ has a subterm of the form $t_1\times(t_2\times t_3)$ or  $(t_1\times t_2)\times t_3$ for some $t_1,t_2,t_3 \in  \rmST(X)$.\hfill$\Box$
\end{compactitem}
\end{definition}

The next observation is obvious by the definition of a large element.

\begin{observation}\label{lm:p-q-r-large}\label{lm:p-generatum-large}\rm $\,$ Let  $p, q, r \in \rmS(X) \setminus \{ 0_\sfS, 1_\sfS \}$.
\begin{compactitem}
\item[(a)] Then  $p \times_{\sfS} (q \times_{\sfS} r)$  is large. 
\item[(b)] If  $p$  is large, then  $p +_{\sfS} q$, $p \times_{\sfS} q$,  and  $q \times_{\sfS} p$  are also large. 
\end{compactitem}
\end{observation}

Next we define a binary relation $\sim_\mathrm{la}$  on  $\rmS(X)$  as follows: for every $q, r \in \rmS(X)$, we let
$q \sim_\mathrm{la} r$  if and only if  $q = r$  or  both  $q$  and  $r$ are large.

\begin{lemma}\label{lm:simL-congruence}\rm The relation $\sim_\mathrm{la}$ is a congruence on $\sfS(X)$.
\end{lemma}
\begin{proof} It is obvious that  $\sim_\mathrm{la}$ is an equivalence relation.
Let $p, q, r \in \rmS(X)$. If $q \sim_\mathrm{la} r$, then by Observation \ref{lm:p-generatum-large}(b), we have
$p +_{\sfS} q \sim_\mathrm{la} p +_{\sfS} r$, $p \times_{\sfS} q \sim_\mathrm{la} p \times_{\sfS} r$, and  $q \times_{\sfS} p \sim_\mathrm{la} r \times_{\sfS} p$.
This proves that $\sim_\mathrm{la}$ is a congruence.
\end{proof}

Then we consider the quotient algebra 
\[ \sfS(X)/_{\sim_\mathrm{la}}=(\rmS(X)/_{\sim_\mathrm{la}}, +_{\sfS}/_{\sim_\mathrm{la}}, \times_{\sfS}/_{\sim_\mathrm{la}}, [0_\sfS]_{\sim_\mathrm{la}}, [1_\sfS]_{\sim_\mathrm{la}}) \]
of $\sfS(X)$ with respect to $\sim_\mathrm{la}$, 
and we abbreviate the above notation by $\sfM(X) =(\rmM(X),\oplus,\otimes,\0,\1)$. Moreover, for  each $q\in \rmS(X)$, we abbreviate $[q]_{\sim_\mathrm{la}}$ by $[q]_\mathrm{la}$.

The algebra $\sfM(X)$ is an almost idempotent  right-distributive strong bimonoid because $\sfS(X)$ is a strong bimonoid which satisfies the same conditions (cf. Lemma \ref{prop:SB-strong-bimonoid}),
and by Lemma \ref{identitiestofactoralgebras} all identities satisfied by $\sfS(X)$ are satisfied also by $\sfM(X)$.

Now we can prove the first main result of this paper.

\begin{theorem}\label{thm:M-is-what-we-want} The strong bimonoid $\sfM(X) =(\rmM(X),\oplus,\otimes,\0,\1)$ is almost idempotent,  right-distributive, weakly locally finite, and not locally finite.
\end{theorem}
\begin{proof} By Observation \ref{obs:biloc-fin+right-disrt-loc-fin}(c),  $\sfM(X)$  is additively locally finite.  

We show that it  is also multiplicatively locally finite as follows.
Fix any  $q \in \rmS(X) \setminus \{ 0_\sfS, 1_\sfS\}$, and let   $p', q', r' \in \rmS(X) \setminus \{ 0_\sfS, 1_\sfS\}$.
Then  $p' \times_{\sfS} (q' \times_{\sfS} r') \sim_\mathrm{la} q \times_{\sfS} (q \times_{\sfS} q)$ because, by Observation \ref{lm:p-q-r-large}(a), both products are large. 
It follows that if  $F$  is a finite subset of $\rmM(X)$,
then the multiplicative submonoid  $\langle F\rangle_{\{\otimes,\1\}}$  of $(\rmM(X), \otimes, \1)$  generated by  $F$ contains $F \cup \{ \1\}$,
all binary products of elements of  $F$  and, possibly, $[q]_\mathrm{la}\otimes ([q]_\mathrm{la}\otimes [q]_\mathrm{la})$.
Thus $\langle F\rangle_{\{\otimes,\1\}}$ is finite.

Hence, $\sfM(X)$  is bi-locally finite. Since  $\sfM(X)$  is right-distributive, by Observation \ref{obs:biloc-fin+right-disrt-loc-fin}(b), it  is weakly locally finite.

It remains to show that  $\sfM(X)$ is not locally finite. 
For this, choose an $x \in X$. We define, for each  $n \in \mathbb{N}$,  the term  $t_n \in \rmST(X)$  by induction as follows:
\begin{align*}
\text{$t_0 = x$  and  $t_{n+1} = x \times (1 + t_n)$.}
\end{align*}
So, e.g.,  $t_1 = x \times (1 + x)$  and $t_2 = x \times (1 + t_1) =  x \times (1 + (x \times (1 + x)))$.

Next, for each $n \in \mathbb{N}$, we consider the $\approx_{E}$-class which contains the term $t_n$, i.e., we
let  $p_n = [t_n]_{E}$. Then we have
\begin{align*}
\text{$p_0 = [x]_{E}$  and  $p_{n+1} = [x]_{E} \times_\sfS (1_\sfS +_\sfS p_n)$  for  each $n \in \mathbb{N}$,}
\end{align*}
hence  $p_n \in \rmS(X)$.

Now we show that, for each  $n \in \mathbb{N}$,  $p_n$  is not large, i.e., there does not exists $t\in [t_n]_{E}$ such that $t$ has the property described in 
Definition \ref{def:large-polynomial}(b). For this, let $t\in [t_n]_{E}$.
By Lemma \ref{lm:approx-characterization} (with $\sfA=\sfST_\Sigma(X)$), we have $t_n \Leftrightarrow^*_{E} t$, i.e., $t_n$ can be transformed to $t$ in finitely many reduction steps  using in each step an identity in $E$ or its inverse.
However, due to the special shape of $t_n$, in each reduction step of the transformation only identity $e_2$ can be used. Hence $t$ cannot be in the form described in Definition \ref{def:large-polynomial}(b), which means that  $p_n$  is not large.

It also follows that the congruence class $p_n$ contains  $2^n$  elements, because there are $n$ occurrences of $+$ in $t_n$, and we can apply the identity $e_2$ in $n$ instances that are independent; thus, there are $2^n$ different elements in the class $p_n$. Consequently, we also obtain that for every $m,n \in \mathbb{N}$ with $m \neq n$, we have $p_m \neq p_n$.

Now for each  $n \in \mathbb{N}$, we let $a_n = [p_n]_{\mathrm{la}} \in \rmM(X)$. Then
$a_{n+1} = [[x]_{E}]_\mathrm{la} \otimes (\1 \oplus  a_n)$  for each  $n \in \mathbb{N}$,
so  $\{a_n \mid n\in \mathbb{N}\} \subseteq \langle \{ \1, a_0 \} \rangle_{\{\oplus, \otimes\}}$.

Finally, let  $m, n \in \mathbb{N}$  with  $m \neq n$. Since  also both  $p_m$  and  $p_n$ are not large, we have $p_m \not\sim_L p_n$, showing  $a_m \neq a_n$,
i.e., that the set $\{a_n \mid n\in \mathbb{N}\}$ is infinite.

Consequently,  $\langle \{ \1, a_0 \} \rangle_{\{\oplus, \otimes\}}$  is infinite,
showing that  $\sfM(X)$  is not locally finite. 
\end{proof}

\section{Weighted tree automata}
\label{sect:wta-def}

\begin{quote}\emph{In this section, let $\Sigma$ be an arbitrary ranked alphabet and  $\B=(B,\oplus,\otimes,\0,\1)$ be an arbitrary  strong bimonoid, if not stated otherwise.}
\end{quote}

A \emph{$(\Sigma,\B)$-weighted tree automaton} (for short: $(\Sigma,\B)$-wta, or simply: wta) is a tuple $\cA=(Q,\delta,F)$ 
 where $Q$ is a finite non-empty set (\emph{states}),
 $\delta=(\delta_k\mid k\in\mathbb{N})$ is a family of mappings $\delta_k: Q^k\times \Sigma^{(k)}\times Q \to B$ (\emph{transition mappings})  where we consider $Q^k$ as set of words over $Q$ of length $k$, and 
 $F: Q \rightarrow B$ is a mapping (\emph{root weight vector}). We denote by $\mathrm{wts}(\cA)$ the set of all weights occurring in $\cA$, i.e., $\mathrm{wts}(\cA) = \bigcup_{k \in \mathbb{N}} \im(\delta_k) \cup \im(F)$.

 Let  $\cA=(Q,\delta,F)$ be a $(\Sigma,\B)$-wta. The {\em vector algebra of $\cA$} is the $\Sigma$-algebra $\V(\cA)=(B^Q,\delta_\cA)$ where, for every $k \in \mathbb{N}$, $\sigma \in \Sigma^{(k)}$, the $k$-ary operation $\delta_\cA(\sigma): B^Q \times \cdots \times B^Q \to B^Q$ is defined by
\begin{equation}\label{eq:delta-A-definition}
\delta_\cA(\sigma)(v_1,\dots,v_k)_q 
  = \bigoplus_{q_1\cdots q_k \in Q^k} \Big(\bigotimes_{i\in[k]} (v_i)_{q_i}\Big) \otimes \delta_k(q_1\cdots q_k,\sigma,q)
  \end{equation}
for every $v_1,\dots,v_k \in B^Q$ and $q \in Q$.  We note that, for each $\alpha \in \Sigma^{(0)}$, we have $\delta_\cA(\alpha)()_q = \delta_0(\varepsilon,\alpha,q)$, because $Q^0=\{\varepsilon\}$ and $\bigotimes_{i\in \emptyset} (v_i)_{q_i}=\1$.

Since the $\Sigma$-term algebra $\sfT_\Sigma$ is initial, there exists a unique $\Sigma$-algebra homomorphism from $\sfT_\Sigma$ to the vector algebra $\V(\cA)$. We denote this homomorphism by $\h_\cA$. Then, for every $t = \sigma(t_1,\ldots,t_k)$ in $\T_\Sigma$ and $q\in Q$, we have
\begin{align*}
\h_\cA(\sigma(t_1,\ldots,t_k))_q &= \h_\cA(\theta_\Sigma(\sigma)(t_1,\ldots,t_k))_q = \delta_{\cA}(\sigma)
(\h_\cA(t_1),\ldots, \h_\cA(t_k))_q \\
&= \bigoplus_{q_1 \cdots q_k \in Q^k} \Big( \bigotimes_{i\in [k]} \h_\cA(t_i)_{q_i}\Big) \otimes \delta_k(q_1\cdots q_k,\sigma,q),
\end{align*}
where $\theta_\Sigma(\sigma)$ is the operation of the $\Sigma$-term algebra associated to $\sigma$; the second equality holds, because $\h_\cA$ is a $\Sigma$-algebra homomorphism. In particular, for each $\alpha \in \Sigma^{(0)}$, we have $\h_\cA(\alpha)_q = \delta_0(\varepsilon,\alpha,q)$.

The \emph{initial algebra semantics of $\cA$}, denoted by $\initialsem{\cA}$, is the weighted tree language $\initialsem{\cA}: \T_\Sigma \rightarrow B$  defined for every $t\in \T_\Sigma$ by  
\begin{equation*}
  \initialsem{\cA}(t) = \bigoplus_{q \in Q} \h_\cA(t)_q \otimes F_q\enspace. 
\end{equation*}

Next we recall the run semantics of the $(\Sigma,\B)$-wta $\cA$.
Let $t \in \T_\Sigma$. We define the \emph{set of positions of $t$}, denoted by $\pos(t)$, by structural induction as follows: (i) For each $t \in \Sigma^{(0)}$, we let $\pos(t) =\{\varepsilon\}$ and (ii) for every $k \in \mathbb{N}_+$, $\sigma \in \Sigma^{(k)}$, and $t_1,\ldots,t_k \in \T_\Sigma$, we let $\pos(\sigma(t_1,\ldots,t_k)) = \{\varepsilon\} \cup \bigcup_{i \in [k]} \{iw \mid w \in \pos(t_i)\}$. In particular, $\pos(t) \subseteq (\mathbb{N}_+)^*$.

Then, for every $t = \sigma(t_1,\ldots,t_k)$ in $\T_\Sigma$, a~\emph{run of $\cA$ on $t$} is a mapping $\rho: \pos(t) \rightarrow Q$. The \emph{set of all runs of $\cA$ on $t$} is denoted by $\R_\cA(t)$. Next we define the mapping $\wt_\cA: \mathrm{TR} \to B$ by structural induction on \(\mathrm{TR} = \{(t,\rho) \mid t \in \T_\Sigma, \rho \in \R_\cA(t)\}\), for every $t = \sigma(t_1,\ldots,t_k)$ in $\T_\Sigma$ and $\rho \in \R_\cA(t)$, as follows: 
 \begin{equation}\label{equ:weight-of-run}
\wt_\cA(t,\rho) = \Big( \bigotimes_{i\in [k]} \wt_\cA(t_i,\rho_i)\Big) \otimes \delta_k\big(\rho(1) \cdots \rho(k),\sigma,\rho(\varepsilon)\big) \enspace,
\end{equation}
where, for each $i \in [k]$, the run $\rho_i: \pos(t_i) \to Q$ of $\cA$ on $t_i$ is defined, for each $w \in \pos(t_i)$, by $\rho_i(w) = \rho(iw)$.

The {\it run semantics of $\cA$}, denoted by $\runsem{\cA}$, is the weighted tree language $\runsem{\cA}:~\T_\Sigma~\rightarrow~B$ such that,  for each $t \in \T_\Sigma$, we let
\begin{equation*}
  \runsem{\cA}(t) = \bigoplus_{\rho \in \R_\cA(t)}\wt(t,\rho) \otimes F_{\rho(\varepsilon)}\enspace. 
\end{equation*}

In general, the initial algebra semantics of $\cA$ is different from the run semantics of $\cA$, cf. e.g., \cite[Ex.~5.2.2-5.2.4]{fulvog22} and also Example~\ref{ex:wta-string-ranked-Trunc}. However, the following equivalence is known (cf. \cite[Thm.~5.3.2]{fulvog22}).

\begin{theorem}{\rm (cf. \cite[Thm.~4.1]{rad10} and \cite[Lm. 4.1.13]{bor04b})} \label{thm:semiring-run=initial} Let $\Sigma$ be a ranked alphabet. Moreover, let $\B=(B,\oplus,\otimes,\0,\1)$ be a strong bimonoid. The following two statements are equivalent:
\begin{compactenum}
\item[(A)] If $\Sigma\not=\Sigma^{(0)}$, then $\B$ is right-distributive, and if $\Sigma\not= \Sigma^{(0)} \cup \Sigma^{(1)}$, then $\B$ is left-distributive.
\item[(B)] For each $(\Sigma,\B)$-wta $\cA$ we have $\runsem{\cA} = \initialsem{\cA}$.
\end{compactenum}
Hence, in particular, if $\B$ is a semiring, then $\initialsem{\cA}=\runsem{\cA}$ for each $(\Sigma,\B)$-wta $\cA$.
\end{theorem}

Essentially, a weighted automaton over words in $\Gamma^*$ (for an alphabet $\Gamma$) is a wta over the string ranked alphabet $\Gamma_e = \Gamma_e^{(0)} \cup \Gamma_e^{(1)}$ where $\Gamma_e^{(0)}=\{e\}$ for some $e \not\in \Gamma$, and $\Gamma_e^{(1)}= \Gamma$. The vector $(\delta(\varepsilon,e,q) \mid q \in Q)$ forms the initial weight vector. For a detailed explanation we refer to \cite[p.~324]{fulvog09new} and \cite[Sec.~3.3]{fulvog22}. We finish this section with an example of a wta over a string ranked alphabet.

 \begin{example}\label{ex:wta-string-ranked-Trunc} \rm 
        We let $\Sigma = \{\gamma_1^{(1)}, \gamma_2^{(1)}, \alpha^{(0)}\}$ and consider the $(\Sigma,\Trunc_{\frac{1}{4}})$-wta $\cA=(Q,\delta,F)$ where $Q=\{q_v,q_1\}$, $F_{q_v}=1$ and $F_{q_1}=0$, and $\delta_0(\varepsilon,\alpha,q_v) = \frac{1}{2}$, $\delta_0(\varepsilon,\alpha,q_1) = 1$, and
        
      $\delta_1(p_1,\gamma_1,p_2) =
      \begin{cases}
        \frac{1}{2} & \text{ if $p_1p_2 =q_vq_v$ }\\
        1 & \text{ if $p_1p_2 =q_1q_1$}\\
        0 & \text{ otherwise}
      \end{cases}$
\ \ \ \  and \ \ \ \ 
       $\delta_1(p_1,\gamma_2,p_2) =
      \begin{cases}
        1 & \text{ if $p_1p_2 =q_vq_v$ }\\
             \frac{1}{2} & \text{ if $p_1p_2 =q_1q_v$}\\
           1 & \text{ if $p_1p_2 =q_1q_1$}\\
        0 & \text{ otherwise}
      \end{cases}$

      For each $n \in \mathbb{N}$, we abbreviate $\gamma_2(\gamma_1(\ldots \gamma_2(\gamma_1(\alpha)) \ldots))$ with $n$ occurrences of $\gamma_2(\gamma_1($ by $(\gamma_2\gamma_1)^n\alpha$. Moreover, we abbreviate $\gamma_1((\gamma_2\gamma_1)^n\alpha)$ by  $\gamma_1(\gamma_2\gamma_1)^n\alpha$.  In particular, $(\gamma_2\gamma_1)^0(\alpha) = \alpha$ and $\gamma_1(\gamma_2\gamma_1)^0(\alpha) = \gamma_1(\alpha)$.

      We will prove that, for each $n \in \mathbb{N}$,
  $\initialsem{\cA}((\gamma_2\gamma_1)^n\alpha) =  b_{2n}$ where $(b_n \mid n \in \mathbb{N})$ is the sequence defined in Example~\ref{ex:Trunc}.      

First, it is easy to see that
      \begin{equation}\label{equ:mon-Trunc-inf-1}
        \text{for each $t \in \T_\Sigma$, we have
          $\h_\cA(t)_{q_1} = 1$}\enspace.
       \end{equation}

      By induction we  prove that the following statement holds.
      \begin{equation}\label{equ:mon-Trunc-inf}
        \text{For each $n \in \mathbb{N}$, we have
          $\h_\cA((\gamma_2\gamma_1)^n\alpha)_{q_v} = b_{2n}$ and $\h_\cA(\gamma_1(\gamma_2\gamma_1)^n\alpha)_{q_v} = b_{2n+1}$}
      \end{equation}

      If $i=0$, then $\h_\cA(\alpha)_{q_v}= \delta_0(\varepsilon,\alpha,q_v)= \frac{1}{2} = b_0$ and
      \[\h_\cA(\gamma_1(\alpha))_{q_v}= \bigoplus_{p\in Q} \h_\cA(\alpha)_p \odot \delta_1(p,\gamma_1,q_v) = \h_\cA(\alpha)_{q_v} \odot \delta_1(q_v,\gamma_1,q_v) = b_0 \cdot \frac{1}{2} = b_1 \enspace,\]
      where the second equality holds because $\delta_1(q_1,\gamma_1,q_v)=0$.
      
      Next we prove the induction step: 
          \begingroup
      \allowdisplaybreaks
      \begin{align*}
        \h_\cA((\gamma_2\gamma_1)^{n+1}\alpha)_{q_v} &= \h_\cA(\gamma_2(\gamma_1((\gamma_2\gamma_1)^n\alpha)))_{q_v}
                                                       = \bigoplus_{p \in Q} \h_\cA(\gamma_1(\gamma_2\gamma_1)^n\alpha)_p \odot \delta_1(p,\gamma_2,q_v)\\
        &= \Big(\h_\cA(\gamma_1(\gamma_2\gamma_1)^n\alpha)_{q_v} \odot \delta_1(q_v,\gamma_2,q_v)\Big) \oplus
          \Big(\h_\cA(\gamma_1(\gamma_2\gamma_1)^n\alpha)_{q_1} \odot \delta_1(q_1,\gamma_2,q_v)\Big)\\
                                                     &= (b_{2n+1} \odot 1) \oplus (1 \odot \frac{1}{2})
                                                       \tag{by induction hypothesis, \eqref{equ:mon-Trunc-inf-1}, and definition of $\delta_1(\ldots,\gamma_2,\ldots)$}\\
                                                     &= b_{2n+1} +  \frac{1}{2}
                                                       \tag{by \eqref{equ:bn-even-odd} }\\
        &= b_{2(n+1)} \enspace.
      \end{align*}
      \endgroup

          \begingroup
      \allowdisplaybreaks
      \begin{align*}
        \h_\cA(\gamma_1(\gamma_2\gamma_1)^{n+1}\alpha)_{q_v} &= \bigoplus_{p \in Q} \h_\cA((\gamma_2\gamma_1)^n\alpha)_p \odot \delta_1(p,\gamma_1,q_v)\\
                                                             &= \h_\cA((\gamma_2\gamma_1)^n\alpha)_{q_v} \odot \delta_1(q_v,\gamma_1,q_v)
        \tag{because $\delta_1(q_1,\gamma_1,q_v)=0)$}\\
                                                     &= b_{2n} \odot \frac{1}{2}
                                                       \tag{by induction hypothesis and definition of $\delta_1(\ldots,\gamma_1,\ldots)$}\\
                                                     &= b_{2n} \cdot \frac{1}{2}
                                                       \tag{by \eqref{equ:bn-even-odd} }\\
        &= b_{2n+1} \enspace.
      \end{align*}
      \endgroup
This finishes the proof of \eqref{equ:mon-Trunc-inf}.

Finally, we can calculate as follows. Let $n \in \mathbb{N}$.
          \begingroup
          \allowdisplaybreaks
          \begin{align*}
              \initialsem{\cA}((\gamma_2\gamma_1)^n\alpha) &=\bigoplus_{p \in Q} \h_\cA((\gamma_2\gamma_1)^n\alpha)_p \odot F_p\\
                                                           &=         \h_\cA((\gamma_2\gamma_1)^n\alpha)_{q_v} \odot F_{q_v}
                                                             \tag{because $F_{q_1}=0$} \\
             &=         b_{2n}  \tag{by \eqref{equ:mon-Trunc-inf} and because $F_{q_v}=1$}\enspace.
          \end{align*}
              \endgroup
              Thus $\{b_{2n} \mid n \in \mathbb{N}\} \subseteq \im(\initialsem{\cA})$. Since $b_i \not= b_j$ for every $i,j \in \mathrm{N}$ with $i\not=j$, the set $\im(\initialsem{\cA})$ is not finite.

In fact, $\initialsem{\cA} \not= \runsem{\cA}$ which can be seen by considering the input tree $t=(\gamma_2\gamma_1)^2\alpha$. Since $F_{q_1}=0$, we only have to consider runs $\rho$ on $t$ such that $\rho(\varepsilon)=q_v$. Among these runs, there are only three runs with weight different from $0$:
\begin{compactitem}
\item $\rho_1$: for each position $w \in \pos(t)$, $\rho_1(w)=q_v$,
\item $\rho_2$: $\rho_2(1111)=\rho_2(111) = q_1$ and $\rho_2(11)=\rho_2(1) = \rho_2(\varepsilon)=q_v$, and 
\item $\rho_3$: $\rho_3(1111)=\rho_3(111) = \rho_3(11)=\rho_3(1) = q_1$ and $\rho_3(\varepsilon)=q_v$. 
\end{compactitem}
Thus we can compute the run semantics of $\cA$ on $t$ as follows:
\begingroup
\allowdisplaybreaks
\begin{align*}
  \runsem{\cA}(t) &= \bigoplus_{\rho \in \R_\cA(t)} \wt(t,\rho) \odot F_{\rho(\varepsilon)}
                    = \bigoplus_{\substack{\rho \in \R_\cA(t):\\\rho(\varepsilon)=q_v}} \wt(t,\rho)\\
                  &= \wt(t,\rho_1) \oplus \wt(t,\rho_2) \oplus \wt(t,\rho_3)\\
                  &= \Big(\frac{1}{2} \odot \frac{1}{2}  \odot 1 \odot \frac{1}{2} \odot 1\Big)
                    \oplus \Big(1 \odot  1 \odot \frac{1}{2} \odot \frac{1}{2}  \odot 1\Big)
                    \oplus \Big(1 \odot 1 \odot 1 \odot 1 \odot \frac{1}{2}\Big)\\
  &= 0 \oplus \frac{1}{4} \oplus \frac{1}{2} = \frac{3}{4}\enspace.
\end{align*}
\endgroup
On the other hand, $\initialsem{\cA}(t) = b_4 = \frac{7}{8}$. Hence $\initialsem{\cA} \not= \runsem{\cA}$.
\hfill $\Box$
  \end{example}


\section{Universality property of initial algebra semantics }\label{sect:wta-section}
 
In this section we will prove Theorem \ref{thm:main-result-on-Sigma}. That is, if the ranked alphabet $\Sigma$ contains a binary symbol, then for each finite subset $A\subseteq B$, we can construct a $(\Sigma,\B)$-wta $\cA$ such that the image of its initial algebra semantics is equal to the closure of $A$, i.e., $\im(\initialsem{\cA}) = \langle A \rangle_{\{\oplus,\otimes,\0,\1\}}$ (cf. Theorem \ref{lm:closure-of-finite-set-i-recognizable}). Weaker versions of this result were proved in \cite[Lm.~6.1]{rad10} (under the assumption that $\Sigma$ contains at least $|A\cup\{\0,\1\}|$ many nullary symbols and at least two binary symbols) and also in \cite[Lm.~12]{fulvog23} (there $\B$ is a bounded lattice and idempotency and commutativity is used in the proof). 

Theorem~\ref{lm:closure-of-finite-set-i-recognizable}, together with Theorem~\ref{thm:M-is-what-we-want},  will be used to prove the main consequence of this paper: there exists a weakly locally finite strong bimonoid $\B$ such that, for each ranked alphabet  $\Sigma$ containing a binary symbol,  there exists a $(\Sigma,\B)$-wta $\cA$ for which $\im(\initialsem{\cA})$ is an infinite set (cf. Theorem \ref{cor:second-main}).

Next we make some preparation for the proof of Theorem~\ref{lm:closure-of-finite-set-i-recognizable}. We assume that $\Sigma$ contains a binary symbol and that $A \cup \{\0,\1\} = \{a_1,\ldots,a_n\}$. Roughly speaking,  we represent the values in $\langle A \rangle_{\{\oplus,\otimes,\0,\1\}}$ by trees over the ranked alphabet $\Delta =\{\oplus^{(2)}, \otimes^{(2)}\} \cup \{a_1^{(0)},\ldots,a_n^{(0)}\}$. 
Formally, we define the mapping $\mathrm{eval}: \T_\Delta \to \langle A \rangle_{\{\oplus,\otimes,\0,\1\}}$ by structural induction, for each $t \in \T_\Delta$, as follows:
\[
  \mathrm{eval}(t) = \begin{cases} a_i & \text{ if $\xi = a_i$ for some $i \in [n]$}\\
    \mathrm{eval}(t_1) \oplus \mathrm{eval}(t_2) & \text{ if $t = \oplus(t_1,t_2)$ for some $t_1,t_2 \in \T_\Delta$}\\
    \mathrm{eval}(t_1) \otimes \mathrm{eval}(t_2) & \text{ if $t = \otimes(t_1,t_2)$ for some $t_1,t_2 \in \T_\Delta$}\enspace.
    \end{cases}
  \]
  Clearly, $\mathrm{eval}$ is surjective. Thus, each value in $\langle A \rangle_{\{\oplus,\otimes,\0,\1\}}$ is represented by at least one tree in $\T_\Delta$.
  
Next we fix arbitrary $\alpha \in \Sigma^{(0)}$ and  $\sigma \in \Sigma^{(2)}$. For each $i \in [0,n]$ we define the $\Sigma$-tree $f_i$ by $f_0 = \alpha$ and $f_{i+1} = \sigma(f_i,\alpha)$ for each $i \in [n-1]$. Then, by structural induction, we define the mapping $g: \T_\Sigma \to \T_\Delta$, for each $t\in \T_\Sigma$, as follows:
\begin{compactitem}
\item if $t=a_i$ for some $i\in[n]$, then $g(t)=f_i$,
\item if $t=\oplus(t_1,t_2)$, then $g(t)=\sigma(\alpha,\sigma(g(t_1),g(t_2)))$, and
\item  if $t=\otimes(t_1,t_2)$, then $g(t)= \sigma(\sigma(g(t_1),g(t_2)),\alpha)$.
\end{compactitem}
In fact, $g$ is a $(\Delta,\Sigma)$-tree homomorphism in the sense of \cite[Def.~3.62]{eng75-15},
and we can consider it as a coding. 

Then we construct the $(\Sigma,\B)$-wta $\cA$ such that $\initialsem{\cA}(g(t)) = \mathrm{eval}(t)$ for each $t \in \T_\Delta$. Thus, since   $\mathrm{eval}$ is surjective, each element of $\langle A \rangle_{\{\oplus,\otimes,\0,\1\}}$ occurs in the image of $\initialsem{\cA}$.

\begin{theorem}\label{lm:closure-of-finite-set-i-recognizable}  Let $\Sigma$ be a ranked alphabet which contains a binary symbol. Moreover, let $\B=(B,\oplus,\otimes,\0,\1)$ be a strong bimonoid and $A \subseteq B$  be a finite subset. Then we can construct a  $(\Sigma,\B)$-wta  $\cA$  such that  $\im(\initialsem{\cA}) = \langle A \rangle_{\{\oplus,\otimes,\0,\1\}}$. In particular, if  $\B$  is generated by  $A$, then we obtain  $\im(\initialsem{\cA}) = B$.
\end{theorem}

\begin{proof} Clearly, $\langle A \rangle_{\{\oplus,\otimes,\0,\1\}} = \langle A \cup \{\0,\1\} \rangle_{\{\oplus,\otimes\}}$. Let $a_1,\ldots,a_n$ be the elements of $A \cup \{\0,\1\}$, i.e., $A \cup \{\0,\1\} = \{a_1,\ldots,a_n\}$. Let $\alpha$ and $\sigma$ be arbitrary elements of $\Sigma^{(0)}$ and $\Sigma^{(2)}$, respectively.
  
  Now we construct the $(\Sigma,\B)$-wta $\cA=(Q,\delta,F)$ as follows (cf. Figure \ref{fig:closure-by-initsem}).
\begin{compactitem}
\item $Q = \{v\} \cup \{q_0,\ldots,q_{n-1}\} \cup \{q_{\mathrm{one}},q_{\oplus}\} \cup \{q_{\otimes}\}$; the intention of the states is as follows:
  \begin{compactitem}
  \item $v$ is the ``main'' state with $\h_\cA(g(t))_v = \mathrm{eval}(t)$ for each $t \in \T_\Delta$ (cf. \eqref{eq:wta-computes-value-on-representations});
  \item the state  $q_i$ with $i \in [0,n-1]$ is used to recognize the tree $f_{i}$ with weight $\1$ (cf. \eqref{eq:usual-values-1}), and in combination with $v$, we will have $\h_\cA(\sigma(f_{i-1},\alpha))_v = a_i$;
  \item the states $q_{\oplus}$ and $q_{\otimes}$ are intermediate states such that $\h_\cA(\sigma(t_1,t_2)))_\oplus = \mathrm{eval}(t_1) \oplus \mathrm{eval}(t_2)$ and  $\h_\cA(\sigma(t_1,t_2)))_\otimes = \mathrm{eval}(t_1) \otimes \mathrm{eval}(t_2)$, respectively; the state $q_{\mathrm{one}}$ supports $q_\oplus$;
    \item the switch from $q_\oplus$ to $v$ and from $q_\otimes$ to $v$ is triggered by the patterns $\sigma(\alpha,.)$ and $\sigma(.,\alpha)$, respectively,
    \end{compactitem}
\item $F_v=\mathbb{1}$ and, for each $q \in Q \setminus \{v\}$, we let $F_q=\mathbb{0}$, and
\item for each $q \in Q$, we define
  \[
    \delta_0(\varepsilon,\alpha,q) =
    \begin{cases}
      \1 & \text{if } q \in \{q_0,q_{\mathrm{one}}\}\\
      \0 & \text{otherwise}
      \end{cases}
    \]
    and, for every $p,q,r \in Q$, we define
    \[
\delta_2(pq,\sigma,r) = 
\left\{
\begin{array}{ll}
\mathbb{1} & \text{if there exists $i \in [n-1]$ such that } pqr= q_{i-1}q_0q_i\\
  a_i & \text{if there exists $i \in [n]$ such that } pqr= q_{i-1}q_0v\\
  \1 & \text{if } pqr \in \{vq_{\mathrm{one}}q_{\oplus}, q_{\mathrm{one}}vq_{\oplus},q_0q_{\oplus}v\} \\
  \1 & \text{if } pqr \in \{vvq_{\otimes}, q_{\otimes}q_0v\}\\
    \1 & \text{if } pqr = q_{\mathrm{one}}q_{\mathrm{one}}q_{\mathrm{one}}\\
  \0 & \text{otherwise} \enspace,
\end{array}
\right.
\]
\item for every $k\in \mathbb{N}$, $\eta \in \Sigma^{(k)}$ with $\sigma\ne \eta\ne \alpha$, and $q,q_1,\ldots,q_k\in Q$, we define $\delta_k(q_1\ldots q_k,\eta,q)=\0$.
\end{compactitem}

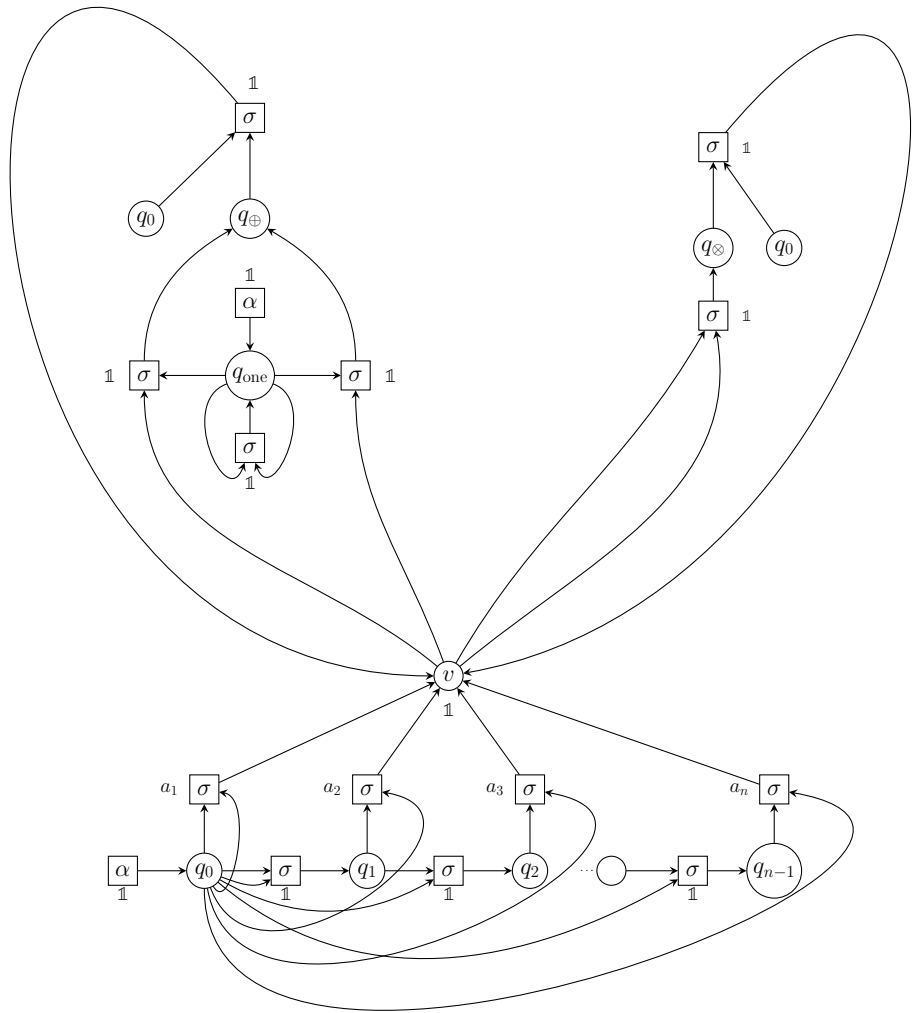
\begin{figure}  
  \centering
  
\begin{tikzpicture}[scale=1.1]
\tikzset{node distance=7em, scale=0.4, transform shape}
\node[state, rectangle] (b) {\huge $\alpha$};
  \node[state, right of=b] (q0) {\huge $q_0$};
\node[state, rectangle, right of=q0](g1-1) {\huge $\sigma$};
  \node[state, right of=g1-1] (q1) {\huge $q_1$};
\node[state, rectangle, right of=q1] (g1-2) {\huge $\sigma$};
  \node[state, right of=g1-2] (q2) {\huge $q_2$};
\node[state, right of=q2,opacity=0] (empty) {}; 
\node[state, rectangle, right of=empty]  (g1-3) {\huge $\sigma$};
\node[state, right of=g1-3] (qn-1) {\huge $q_{n-1}$};

\node[state, rectangle,above of=q0] (h0) {\huge $\sigma$};
\node[state, rectangle,above of=q1] (h1) {\huge $\sigma$};
\node[state, rectangle,above of=q2] (h2) {\huge $\sigma$};
\node[state, rectangle,above of=qn-1] (hn-1) {\huge $\sigma$};

\tikzset{node distance=3em}
\node[left of=h0]     {\LARGE $a_1$};
\node[left of=h1]     {\LARGE $a_2$};
\node[left of=h2]     {\LARGE $a_3$};
\node[left of=hn-1]     {\LARGE $a_n$};

\draw (q0)    edge[->,>=stealth, out=300, in=350] (h0);
\draw (q0)    edge[->,>=stealth, out=90, in=270] (h0);

\draw (q0)    edge[->,>=stealth, out=290, in=350,looseness=1.8] (h1);
\draw (q1)    edge[->,>=stealth, out=90, in=270] (h1);

\draw (q0)    edge[->,>=stealth, out=280, in=350, looseness=1.5] (h2);
\draw (q2)    edge[->,>=stealth, out=90, in=270] (h2);

\draw (q0)    edge[->,>=stealth, out=270, in=350, looseness=1.3] (hn-1);
\draw (qn-1)    edge[->,>=stealth, out=90, in=270] (hn-1);

\node (dots) at ($(q2.east)!0.5!(empty.east)$) {$\ldots$};

\tikzset{node distance=2em}
\node[below of=b]    (wb)    {\LARGE $\1$};
\node[below of=g1-1] (wg1-1) {\LARGE $\1$};
\node[below of=g1-2] (wg1-2) {\LARGE $\1$};
\node[below of=g1-3] (wg1-3) {\LARGE $\1$};
(
\draw (b)    edge[->,>=stealth] (q0);
\draw (q0)   edge[->,>=stealth, out=0, in=180] (g1-1);
\draw (q0)   edge[->,>=stealth, out=-20, in=210] (g1-1);
\draw (g1-1) edge[->,>=stealth] (q1);
\draw (q1)   edge[->,>=stealth, out=0, in=180] (g1-2);
\draw (q0)   edge[->,>=stealth, out=-30, in=210] (g1-2);
\draw (g1-2) edge[->,>=stealth] (q2);
\draw (empty)edge[->,>=stealth] (g1-3);
\draw (g1-3) edge[->,>=stealth] (qn-1);
\draw (q0)   edge[->,>=stealth, out=-40, in=210] (g1-3);

\node[state, above=5cm of g1-2] (v) {\huge $v$};
\draw (h0)    edge[->,>=stealth] (v);
\draw (h1)    edge[->,>=stealth] (v);
\draw (h2)    edge[->,>=stealth] (v);
\draw (hn-1)    edge[->,>=stealth] (v);
\node[below of=v, yshift=-3mm]     {\LARGE $\1$};

\node[state, rectangle,above=6cm of v,xshift=-6cm] (sigma1) {\huge $\sigma$};
\node[state,above=1cm of sigma1] (state1) {\huge $q_{\mathrm{one}}$};
\draw (state1) edge[->,>=stealth,out=200, in=250,looseness=1.7]  (sigma1);
\draw (state1) edge[->,>=stealth,out=340, in=290,looseness=1.6]  (sigma1);
\draw (sigma1) edge[->,>=stealth]  (state1);
\tikzset{node distance=3em}
\node[below of=sigma1]     {\LARGE $\1$};

\node[state, rectangle,left=2cm of state1] (sigmaleftleft) {\huge $\sigma$};
\draw (v) edge[->,>=stealth,out=140, in=270]  (sigmaleftleft);
\draw (state1) edge[->,>=stealth]  (sigmaleftleft);
\node[left of=sigmaleftleft]     {\LARGE $\1$};

\node[state, rectangle,right=2cm of state1] (sigmaleftright) {\huge $\sigma$};
\draw (v) edge[->,>=stealth,out=110, in=270]  (sigmaleftright);
\draw (state1) edge[->,>=stealth]  (sigmaleftright);
\node[right of=sigmaleftright]     {\LARGE $\1$};

\node[state, rectangle,above=1cm of state1] (alpha1) {\huge $\alpha$};
\draw (alpha1) edge[->,>=stealth] (state1);
\node[above of=alpha1,yshift=-2mm]     {\LARGE $\1$};

\node[state,above=1.5cm of alpha1] (oplus) {\huge $q_\oplus$};
\draw (sigmaleftleft) edge[->,>=stealth,out=90,in=210] (oplus);
\draw (sigmaleftright) edge[->,>=stealth,out=90,in=330] (oplus);

\node[state,rectangle,above=2cm of oplus] (sigmatopleft) {\huge $\sigma$};
\draw (sigmatopleft) edge[->,>=stealth,out=130,in=180, looseness=2.2] (v);
\node[state,left=2cm of oplus] (q0left) {\huge $q_0$};
\draw (q0left) edge[->,>=stealth] (sigmatopleft);
\draw (oplus) edge[->,>=stealth] (sigmatopleft);
\node[above of=sigmatopleft, xshift=0.1cm]     {\LARGE $\1$};

\node[state, rectangle,above=10cm of v,xshift=8cm] (sigma2) {\huge $\sigma$};
\node[right=3mm of sigma2] {\Large $\1$};
\node[state,above=1cm of sigma2] (otimes) {\huge $q_\otimes$};
\draw (v) edge[->,>=stealth,out=60,in=240] (sigma2);
\draw (v) edge[->,>=stealth,out=40,in=280] (sigma2);
\draw (sigma2) edge[->,>=stealth] (otimes);
\node[state, rectangle, above=2cm of otimes] (sigmatopright) {\huge $\sigma$}; 
\draw (otimes) edge[->,>=stealth] (sigmatopright);
\draw (sigmatopright) edge[->,>=stealth,out=50,in=10,looseness=2.1] (v);
\node[right=3mm of sigmatopright] {\Large $\1$};
\node[state,right=1cm of otimes] (q0right) {\huge $q_0$};
\draw (q0right) edge[->,>=stealth] (sigmatopright);
\end{tikzpicture}

  \caption{\label{fig:closure-by-initsem} The $(\Sigma,\B)$-wta $\cA$ of the proof of Theorem \ref{lm:closure-of-finite-set-i-recognizable}, where the three occurrences of the state $q_0$ have to be identified. }
\end{figure}

\

Next we prove that $\im(\initialsem{\cA}) = \langle A \rangle_{\{\oplus,\otimes,\0,\1\}}$. Since $\mathrm{wts}(\cA) = A \cup \{\0,\1\}$, we have $\im(\initialsem{\cA}) \subseteq \langle A \rangle_{\{\oplus,\otimes,\0,\1\}}$. Thus it remains to prove $\langle A \rangle_{\{\oplus,\otimes,\0,\1\}}\subseteq \im(\initialsem{\cA})$.

For this purpose, we need some auxiliary statements, which are easy to see.
\begin{equation}
  \text{For each $q \in \{v,q_\oplus,q_\otimes\} \cup \{q_1,\ldots,q_{n-1}\}$, we have $\h_\cA(\alpha)_q=\0$\enspace.} \label{eq:usual-values-3}
    \end{equation}
\begin{equation}
\text{For each $i \in [0,n]$, we have $\h_\cA(f_{i})_{q_\otimes}=\0$ \enspace.}
 \label{eq:usual-values-1.5}
\end{equation}
  \begin{equation}
   \text{For every $s_1,s_2 \in \T_\Sigma$, we have $\h_\cA(\sigma(s_1,s_2))_{q_0} = \0$\enspace.} \label{eq:usual-values-4}
 \end{equation}
  \begin{equation}
   \text{For every $s_1,s_2,s_3 \in \T_\Sigma$ and $i \in [0,n-1]$, we have $\h_\cA(\sigma(s_1,\sigma(s_2,s_3)))_{q_i} = \0$ \enspace.} \label{eq:usual-values-5}
 \end{equation}
 \begin{equation}
   \text{For each $s \in \T_\Sigma$, we have $\h_\cA(s)_{q_{\mathrm{one}}}=\1$\enspace.} \label{eq:usual-values-6}
 \end{equation}

As further preparation, we deal with the values of expressions of the form $\h_\cA(f_i)_{q_j}$. 
First, by bounded induction on $i$, we prove the following:
\begin{equation}
  \text{For each $i,j \in [0,n-1]$ with $i \le j$, we have $\h_\cA(f_i)_{q_{j}}=\h_\cA(f_0)_{q_{j-i}}$ \enspace.} \label{eq:usual-values-1.1}
\end{equation}
The case $i=0$ is trivial. For the induction step, let $i+1 \le j$. We can calculate as follows:
\begin{align*}
\h_\cA(f_{i+1})_{q_{j}} &= \h_\cA(\sigma(f_i,\alpha))_{q_j} = \h_\cA(f_i)_{q_{j-1}} \otimes \h_\cA(\alpha)_{q_0} \otimes \delta_2(q_{j-1}q_0,\sigma,q_j) = \h_\cA(f_i)_{q_{j-1}} = \h_\cA(f_0)_{q_{j-(i+1)}} \enspace,
\end{align*}
where the last equality holds by induction hypothesis. This proves \eqref{eq:usual-values-1.1}.

Second, by bounded induction on $i$, we prove the following:
\begin{equation}
  \text{For each $i,j \in [0,n-1]$ with $i \ge j$, we have $\h_\cA(f_i)_{q_{j}}=\h_\cA(f_{i-j})_{q_0}$ \enspace.} \label{eq:usual-values-1.11}
\end{equation}
The case $i=0$ is trivial again. For the induction step, let $i+1 \ge j$. As above we obtain $\h_\cA(f_{i+1})_{q_{j}} =  \h_\cA(f_i)_{q_{j-1}}$. Then, by induction hypothesis, we obtain that $\h_\cA(f_i)_{q_{j-1}} = \h_\cA(f_{(i+1)-j})_{q_0}$. This  proves \eqref{eq:usual-values-1.11}.

Since,
\begin{compactitem}
\item for each $i < j$, Equality \eqref{eq:usual-values-1.1} implies that $\h_\cA(f_i)_{q_{j}} = \h_\cA(f_0)_{q_{j-i}} = \h_\cA(\alpha)_{q_{j-i}} = \0$ (because $q_{j-i} \not= q_0$), and
\item for each $i > j$, Equality \eqref{eq:usual-values-1.11} implies that $\h_\cA(f_i)_{q_{j}}=\h_\cA(f_{i-j})_{q_0}= \0$ (because $f_{i-j}\not= \alpha$), and
\item for $i=j$, Equalities \eqref{eq:usual-values-1.1} and \eqref{eq:usual-values-1.11} imply that $\h_\cA(f_i)_{q_{j}}=\h_\cA(f_0)_{q_{j-i}} = \h_\cA(f_{i-j})_{q_0}= \h_\cA(f_0)_{q_0} = \h_\cA(\alpha)_{q_0}= \1$,
\end{compactitem}
the following two statements hold.
\begin{equation}
  \text{For each $i \in [0,n-1]$, we have $\h_\cA(f_i)_{q_{i}}  = \1$ \enspace.} \label{eq:usual-values-1}
\end{equation}
\begin{equation}
  \text{For each $i,j \in [0,n-1]$ with $j \not= i$, we have $\h_\cA(f_i)_{q_{j}}=\0$ \enspace.} \label{eq:usual-values-1.2}
\end{equation}
This finishes the proofs of auxiliary statements.

By structural induction we prove the following statement:
\begin{equation}\label{eq:wta-computes-value-on-representations}
  \text{For each $t \in \T_\Delta$, we have $\h_\cA(g(t))_v = \mathrm{eval}(t)$} \enspace,
  \end{equation}
 where the ranked alphabet $\Delta$ and the mappings $\mathrm{eval}: \T_\Delta \to \langle A \rangle_{\{\oplus,\otimes,\0,\1\}}$ 
 and $g:\T_\Delta \to \T_\Sigma $ are defined before this theorem.

First let $t=a_i$ for some $i \in [n]$. Then $g(a_i)=f_i$ and we can calculate as follows.
\begingroup
\allowdisplaybreaks
\begin{align*}
  \h_\cA(g(a_i))_v=& \ \h_\cA(f_i)_v 
                    = \bigoplus_{p_1p_2\in Q^2} \h_\cA(f_{i-1})_{p_1} \otimes \h_\cA(\alpha)_{p_2} \otimes \delta_2(p_1p_2,\sigma,v)\\[2mm]
  = \ & \Big(\bigoplus_{j \in [0,n-1]} \h_\cA(f_{i-1})_{q_j} \otimes \h_\cA(\alpha)_{q_0} \otimes \delta_2(q_jq_0,\sigma,v) \Big)\\
                  & \oplus \Big( \h_\cA(f_{i-1})_{q_0} \otimes \h_\cA(\alpha)_{q_\oplus} \otimes \delta_2(q_0q_\oplus,\sigma,v)\Big)\\
                    & \oplus \Big( \h_\cA(f_{i-1})_{q_\otimes} \otimes \h_\cA(\alpha)_{q_0} \otimes \delta_2(q_\otimes q_0,\sigma,v)\Big)
        \tag{because for each other combination of $p_1p_2$ we have $\delta_2(p_1p_2,\sigma,v) = \0$}\\[2mm]
  = \ & \Big(\bigoplus_{j \in [0,n-1]} \h_\cA(f_{i-1})_{q_j} \otimes \1 \otimes \delta_2(q_jq_0,\sigma,v) \Big)\\
                  & \oplus \Big( \h_\cA(f_{i-1})_{q_0} \otimes \0 \otimes \delta_2(q_0q_\oplus,\sigma,v)\Big)\\
                  & \oplus \Big( \0 \otimes \h_\cA(\alpha)_{q_0} \otimes \delta_2(q_\otimes q_0,\sigma,v)\Big)
                    \tag{by \eqref{eq:usual-values-1}(for $i=0$), \eqref{eq:usual-values-3}, and \eqref{eq:usual-values-1.5} }\\[2mm]
   = \ & \bigoplus_{j \in [0,n-1]} \h_\cA(f_{i-1})_{q_j} \otimes \delta_2(q_jq_0,\sigma,v) \\[2mm]
  = \ & \h_\cA(f_{i-1})_{q_{i-1}} \otimes \delta_2(q_{i-1}q_0,\sigma,v)
  \tag{by \eqref{eq:usual-values-1.2}}\\[2mm]
  = \ & \1 \otimes a_i \tag{by \eqref{eq:usual-values-1} and because $ \delta_2(q_{i-1}q_0,\sigma,v)=a_i$}\\[2mm]
        = \ & \mathrm{eval}(a_i) \enspace.
  \end{align*}
  \endgroup

For the induction step, we distinguish two cases. First, let $t = \oplus(t_1,t_2)$ for some $t_1,t_2 \in \T_\Delta$. Then: 
\begingroup
\allowdisplaybreaks
\begin{align*}
  &\h_\cA\big(g(\oplus(t_1,t_2))\big)_v = \h_\cA\big(\sigma(\alpha,\sigma(g(t_1),g(t_2)))\big)_v \\[2mm]
  = \  & \bigoplus_{p_1p_2\in Q^2} \h_\cA(\alpha)_{p_1} \otimes \h_\cA\big(\sigma(g(t_1),g(t_2))\big)_{p_2} \otimes \delta_2(p_1p_2,\sigma,v)\\[2mm]
  = \ & \Big(\bigoplus_{j \in [0,n-1]} \h_\cA(\alpha)_{q_j} \otimes \h_\cA\big(\sigma(h(t_1),h(t_2))\big)_{q_0} \otimes \delta_2(q_jq_0,\sigma,v) \Big)\\
  & \oplus \Big( \h_\cA(\alpha)_{q_0} \otimes \h_\cA\big(\sigma(g(t_1),g(t_2))\big)_{q_\oplus} \otimes \delta_2(q_0q_\oplus,\sigma,v)\Big)\\
  & \oplus \Big( \h_\cA(\alpha)_{q_\otimes} \otimes \h_\cA\big(\sigma(g(t_1),g(t_2))\big)_{q_0} \otimes \delta_2(q_\otimes q_0,\sigma,v)\Big)
    \tag{because for each other combination $p_1p_2$ we have $\delta_2(p_1p_2,\sigma,v)=\0$}\\[2mm]
   = \ & \Big(\bigoplus_{j \in [0,n-1]} \h_\cA(\alpha)_{q_j} \otimes \0 \otimes \delta_2(q_jq_0,\sigma,v) \Big)\\
  & \oplus \Big( \h_\cA(\alpha)_{q_0} \otimes \h_\cA\big(\sigma(g(t_1),g(t_2))\big)_{q_\oplus} \otimes \delta_2(q_0q_\oplus,\sigma,v)\Big)\\
  & \oplus \Big( \h_\cA(\alpha)_{q_\otimes} \otimes \0 \otimes \delta_2(q_\otimes q_0,\sigma,v)\Big)
  \tag{by \eqref{eq:usual-values-4}}\\[2mm]
  = \ & \h_\cA(\alpha)_{q_0} \otimes \h_\cA\big(\sigma(g(t_1),g(t_2))\big)_{q_\oplus} \otimes \delta_2(q_0q_\oplus,\sigma,v)\\[2mm]
  = \ & \1 \otimes \h_\cA\big(\sigma(g(t_1),g(t_2))\big)_{q_\oplus} \otimes \1
        \tag{by \eqref{eq:usual-values-1} (for $i=0$) and definition of $\delta_2$}\\[2mm]
  = \ &  \h_\cA\big(\sigma(g(t_1),g(t_2))\big)_{q_\oplus} \\[2mm]
  = \ & \Big(\h_\cA(g(t_1))_v \otimes \h_\cA(g(t_2))_{q_{\mathrm{one}}} \otimes \delta_2(q_vq_{\mathrm{one}},\sigma,q_\oplus)\Big) \oplus
        \Big(\h_\cA(g(t_1))_{q_{\mathrm{one}}} \otimes \h_\cA(g(t_2))_v \otimes \delta_2(q_{\mathrm{one}}q_v,\sigma,q_\oplus)\Big)
  \tag{because for each other combination $p_1p_2$ we have $\delta_2(p_1p_2,\sigma,q_\oplus)=\0$}\\[2mm]
  = \ & (\mathrm{eval}(t_1) \otimes \1 \otimes \1) \oplus
        (\1 \otimes \mathrm{eval}(t_2) \otimes \1)
   \tag{by induction hypothesis, \eqref{eq:usual-values-6}, and definition of $\delta_2$}\\[2mm]
  = \ & \mathrm{eval}(t_1) \oplus \mathrm{eval}(t_2) = \mathrm{eval}(\oplus(t_1,t_2))\enspace.
\end{align*}
\endgroup

\

Secondly, let $t = \otimes(t_1,t_2)$ for some $t_1,t_2 \in \T_\Delta$. Then:
\begingroup
\allowdisplaybreaks
\begin{align*}
  &\h_\cA(\otimes(t_1,t_2))_v = \h_\cA\big(\sigma(\sigma(g(t_1),g(t_2)),\alpha)\big)_v \\[2mm]
  = \  & \bigoplus_{p_1p_2\in Q^2} \h_\cA\big(\sigma(g(t_1),g(t_2))\big)_{p_1} \otimes \h_\cA(\alpha)_{p_2} \otimes \delta_2(p_1p_2,\sigma,v)\\
  = \ & \Big(\bigoplus_{j \in [0,n-1]} \h_\cA\big(\sigma(g(t_1),g(t_2))\big)_{q_j} \otimes \h_\cA(\alpha)_{q_0} \otimes \delta_2(q_jq_0,\sigma,v) \Big)\\[2mm]
   & \oplus \Big( \h_\cA\big(\sigma(g(t_1),g(t_2))\big)_{q_0} \otimes \h_\cA(\alpha)_{q_\oplus} \otimes \delta_2(q_0q_\oplus,\sigma,v)\Big)\\
  & \oplus \Big( \h_\cA\big(\sigma(g(t_1),g(t_2))\big)_{q_\otimes} \otimes \h_\cA(\alpha)_{q_0} \otimes \delta_2(q_\otimes q_0,\sigma,v)\Big)
    \tag{because for each other combination $p_1p_2$ we have $\delta_2(p_1p_2,\sigma,v)=\0$}\\[2mm]
   = \ & \Big(\bigoplus_{j \in [0,n-1]} \0 \otimes \h_\cA(\alpha)_{q_0} \otimes \delta_2(q_jq_0,\sigma,v) \Big)\\
  & \oplus \Big( \h_\cA\big(\sigma(g(t_1),g(t_2))\big)_{q_0} \otimes \0 \otimes \delta_2(q_0q_\oplus,\sigma,v)\Big)\\
  & \oplus \Big( \h_\cA\big(\sigma(g(t_1),g(t_2))\big)_{q_\otimes} \otimes \1 \otimes \delta_2(q_\otimes q_0,\sigma,v)\Big) \tag{by \eqref{eq:usual-values-5} using that $g(t_2)(\varepsilon)=\sigma$, \eqref{eq:usual-values-3}, and \eqref{eq:usual-values-1}(for $i=0$)}\\[2mm]
  = \ & \h_\cA\big(\sigma(g(t_1),g(t_2))\big)_{q_\otimes} \otimes \delta_2(q_\otimes q_0,\sigma,v)\\[2mm]
  = \ & \h_\cA\big(\sigma(g(t_1),g(t_2))\big)_{q_\otimes} \tag{by definition of $\delta_2$}\\[2mm]
  = \ & \h_\cA(g(t_1))_v \otimes \h_\cA(g(t_2))_v \otimes \delta_2(vv,\sigma,{q_\otimes})
  \tag{because for each other combination $p_1p_2$ we have $\delta_2(p_1p_2,\sigma,q_\otimes)=\0$}\\[2mm]
  = \ & \mathrm{eval}(t_1) \otimes \mathrm{eval}(t_2) \tag{by induction hypothesis and definition of $\delta_2$}\\[2mm]
= \ & \mathrm{eval}(\otimes(t_1,t_2)) \enspace. 
\end{align*}
\endgroup
This finishes the proof of  \eqref{eq:wta-computes-value-on-representations}.

Now let $a \in \langle A\rangle_{\{\oplus,\otimes,\0,\1\}}$. Since  $\mathrm{eval}$ is surjective, there exists $t \in \T_\Delta$ such that $\mathrm{eval}(t) = a$.  Then
\[
\initialsem{\cA}(g(t)) = \bigoplus_{p \in Q} \h_\cA(g(t))_p \otimes F_p = \h_\cA(g(t))_v = \mathrm{eval}(t) = a
  \]
where the last but one equality follows from   \eqref{eq:wta-computes-value-on-representations}.
Hence $\langle A\rangle_{\{\oplus,\otimes,\0,\1\}} \subseteq \im(\initialsem{\cA})$.
\end{proof}

We note that the proof of Theorem \ref{lm:closure-of-finite-set-i-recognizable}  is effective, even if the strong bimonoid  $\B$  is not given effectively:
Given the ranked alphabet  $\Sigma$  and the subset  $A$  generating  $\B$  as input data, the proof gives 
the construction of the requested wta  $\cA$  satisfying   $\im(\initialsem{\cA}) = \langle A\rangle_{\{\oplus,\otimes,\0,\1\}}$.

We believe that Theorem \ref{lm:closure-of-finite-set-i-recognizable} holds for each ranked alphabet $\Sigma$ which  satisfies that $|\Sigma^{(k)}| \ge 1$ for some $k\ge 2$. As an immediate consequence of Theorem~\ref{lm:closure-of-finite-set-i-recognizable}, we obtain the following result.

  \begin{corollary} \label{lm:weak-loc-fin-not-loc-fin-monadic-is-weaker}\rm Let $\Sigma$ be a ranked alphabet and $\B$ a strong bimonoid.  If $\Sigma$ contains a binary symbol and $\B$ is not locally finite, then there exists a  $(\Sigma,\B)$-wta $\cA$ such that the mapping $\initialsem{\cA}$ has infinite image.
       \end{corollary}

  \begin{proof} Since $\B$ is not locally finite, there exists a finite set $A\subseteq B$ such that  $\langle A \rangle_{\{\oplus,\otimes,\0,\1\}}$ is an infinite set. Since $\Sigma$ contains a binary symbol, by Theorem \ref{lm:closure-of-finite-set-i-recognizable} we can construct a $(\Sigma,\B)$-wta $\cA$ such that $\im(\initialsem{\cA}) = \langle A \rangle_{\{\oplus,\otimes,0,1\}}$.
  \end{proof}

Now it is easy to derive the main consequence of our results.

\begin{theorem}\label{cor:second-main} \rm There exists a weakly locally finite strong bimonoid $\B$ such that for each ranked alphabet $\Sigma$ which contains a binary symbol there exists a $(\Sigma,\B)$-wta $\cA$
for which $\im(\initialsem{\cA})$ is an infinite set.
\end{theorem}
\begin{proof} It follows from Theorem \ref{thm:M-is-what-we-want} and Corollary \ref{lm:weak-loc-fin-not-loc-fin-monadic-is-weaker}.
\end{proof}

Due to the equality of initial algebra semantics and run semantics of wta over semiring we obtain the following consequence of Theorem~\ref{lm:closure-of-finite-set-i-recognizable}.

\begin{corollary}\label{cor:second-main-semirings} \rm Let $\Sigma$ be an arbitrary ranked alphabet containing a binary symbol, and let $\B =(B,\oplus,\otimes,\0,\1)$ be any finitely generated semiring. Then there exists a weighted tree automaton $\cA$ over $\Sigma$ and $\B$ such
  that $\im(\runsem{\cA}) = B$.
  \end{corollary}
  \begin{proof} This is immediate by Theorem~\ref{thm:semiring-run=initial} and Theorem~\ref{lm:closure-of-finite-set-i-recognizable}.  
    \end{proof}


\section{Further research}\label{sect:further-research}

For the classes of locally finite strong bimonoids and bi-locally finite strong bimonoids, we know the following nice characterizations in terms of the  finite image property of initial algebra semantics and of run semantics, respectively. 

\begin{theorem}\label{thm:loc-finite-rec-step-function} (cf. \cite[Thm.~16.1.6]{fulvog22}) Let  $\B$ be a strong bimonoid. Then the following two statements are equivalent.
\begin{compactenum}
\item[(A)]  $\B$ is locally finite.
\item[(B)]  For each ranked alphabet $\Sigma$ and for each $(\Sigma,\B)$-wta $\cA$, the set $\im(\initialsem{\cA})$  is finite.
\end{compactenum}
\end{theorem}

\begin{theorem} \label{thm:bi-loc-finite-rec-step-function} (cf. \cite[Thm.~16.2.7]{fulvog22})  Let $\B$ be a strong bimonoid. Then the following two statements are equivalent.
\begin{compactenum}
\item[(A)]  $\B$ is bi-locally finite.
\item[(B)]  For each ranked alphabet $\Sigma$ and for each $(\Sigma,\B)$-wta $\cA$, the set $\im(\runsem{\cA})$  is finite.
\end{compactenum}
\end{theorem}

For the class of weakly locally finite strong bimonoids, no such characterization is known and we only have the following implications.

\begin{theorem}\label{thm:wlf-implies-fin-im-monadic}   Let $\B$ be a strong bimonoid. Then Statement (A) implies Statement (B), and Statement (B) implies Statement (C).
\begin{compactenum}
\item[(A)]  $\B$ is weakly locally finite.
\item[(B)]  For each monadic ranked alphabet $\Sigma$ and for each $(\Sigma,\B)$-wta $\cA$, the set $\im(\initialsem{\cA})$  is finite.
  \item[(C)] $\B$ is bi-locally finite.
\end{compactenum}
\end{theorem}
\begin{proof} (A) $\Rightarrow$ (B): \cite[Lm.~16.1.1]{fulvog22} and \cite[Lm.~18]{drostuvog10} for string ranked alphabets.

  (B) $\Rightarrow$ (C): First we show that $(B,\oplus,\0)$ is locally finite.  Let  $b \in B$. Consider the string ranked alphabet $\Gamma_e=\{\gamma^{(1)}, e^{(0)}\}$ and the following wta over $\Gamma_e$  and  $\B$ (i.e., essentially a weighted word automaton over $\{\gamma\}$ and $\B$):  $\cA = (Q, \delta,F)$  with  $Q = \{ p, q \}, \delta_0(\varepsilon,e,p) = \delta_0(\varepsilon,e,q) = b$,  $F_p = F_q = \1$, and $\delta_1(r,\gamma,r') = \1$  for  all  $r, r' \in Q$.

For each $n \in \mathbb{N}$, let us put  $nb = b \oplus \ldots \oplus b$ ($n$  summands). Then it is easy to see that, for every $n \in \mathbb{N}$ and $r \in Q$: $\h_\cA(\gamma^n(e))_r = 2^nb$, and hence $\initialsem{\cA}(\gamma^n(e)) = \h_\cA(\gamma^n(e))_p \oplus \h_\cA(\gamma^n(e))_q = 2^{n+1}b$.

By assumption (B), it follows that the cyclic submonoid $(\langle b \rangle_{\{\oplus,\0\}},\oplus,\0)$  is finite.
Since addition is commutative, the monoid $(B,\oplus,\0)$  is locally finite.

Second we show that $(B,\otimes,\1)$ is locally finite. Let  $A \subseteq B$  be a finite non-empty subset.
Consider the string ranked alphabet $\Gamma_e = \{e^{(0)}\} \cup \{b^{(1)} \mid b \in A\}$ and
the following one-state wta over $\Gamma_e$ and  $\B$: $\cA = (Q, \delta,F)$  with  $Q = \{ q \}$, $F_q=\1$, $\delta_0(\varepsilon,e,q) = \1$, and
$\delta_1(q,b,q) = b$  for each  $b \in A$.

Then for each  $w = b_1\ldots b_n \in A^*$  we have
$\initialsem{\cA}(w(e)) = b_1 \otimes \ldots \otimes b_n$.

Hence  $\im(\initialsem{\cA}) = \langle A \rangle_{\{\otimes\}}$. By assumption (B), the set  $\langle A \rangle_{\{\otimes\}}$  is finite. Therefore  $(B,\otimes,\1)$   locally finite. 
  \end{proof}

In \cite[Lm.~18]{drostuvog10}, Theorem~\ref{thm:wlf-implies-fin-im-monadic}(B)$\Rightarrow$(C) was proved with (B) replaced by (B') where
\begin{compactenum}
  \item[(B')] For each string ranked alphabet $\Sigma$ with $|\Sigma^{(1)}| \ge 2$ and each $(\Sigma,\B)$-wta $\cA$, the set $\im(\initialsem{\cA})$  is finite or the set $\im(\runsem{\cA})$  is finite.
\end{compactenum}

It would be desirable to define a class of strong bimonoids, say \emph{boundedly locally finite strong bimonoids} such that the following holds.

\begin{claim}\label{thm:monadic-wlf-char-fin-im-prop-init} \rm Let $\B$ be a strong bimonoid. Then the following two statements are equivalent.
\begin{compactenum}
\item[(A)]  $\B$ is boundedly weakly locally finite.
\item[(B)]  For each monadic ranked alphabet $\Sigma$ and for each $(\Sigma,\B)$-wta $\cA$, the set $\im(\initialsem{\cA})$  is finite.
\end{compactenum}
\end{claim}

If Claim \ref{thm:monadic-wlf-char-fin-im-prop-init} holds, then by using Theorem \ref{thm:wlf-implies-fin-im-monadic}, we have
\[
\text{weakly locally finite \ $\Rightarrow$ \ boundedly locally finite \ $\Rightarrow$ bi-locally finite.}
  \]

The strong bimonoid $\mathsf{Stb}$ in Example \ref{ex:Stb} is bi-locally finite and not boundedly locally finite.
This holds because in \cite[Ex.~25]{drostuvog10} a wta $\cA$ over a string ranked alphabet and $\mathsf{Stb}$ is given such that $\im(\initialsem{\cA})$ is infinite; and since each string ranked alphabet is monadic, by  Claim \ref{thm:monadic-wlf-char-fin-im-prop-init}$\neg$(B) $\Rightarrow$ $\neg$(A), the strong bimonoid $\mathsf{Stb}$ is not boundedly locally finite.

\begin{claim} \rm There exists a boundedly locally finite strong bimonoid which is not weakly locally finite.
\end{claim}
\noindent \emph{Argumentation:}  The class of weakly locally finite strong bimonoids is too restrictive:
\begin{compactitem}
\item it requires finiteness over unbounded summations, whereas in the initial algebra semantics of a wta over a monadic ranked alphabet the number of summands is bounded (in fact, equal) to the number of states and \\
\item  it requires finiteness for the summation of two summands which are generated at different ``levels'' of the closure, whereas in the initial algebra semantics of a wta over a monadic ranked alphabet the summands come from the same level. \hfill $\Box$
\end{compactitem}



\begin{thebibliography}{DGMM11}

\bibitem[Bir93]{bir93}
G.~Birkhoff.
\newblock {\em Lattice Theory}, volume~25 of {\em Colloquium Publications}.
\newblock American Mathematical Society, 1993.
\newblock 3rd edition.

\bibitem[BN98]{baanip98}
F.~Baader and T.~Nipkow.
\newblock {\em Term Rewriting and All That}.
\newblock Cambridge University Press, 1998.

\bibitem[Bor04]{bor04}
B.~Borchardt.
\newblock A pumping lemma and decidability problems for recognizable tree
  series.
\newblock {\em Acta Cybern.}, 16(4):509--544, 2004.

\bibitem[Bor05]{bor04b}
B.~Borchardt.
\newblock {\em The Theory of Recognizable Tree Series}.
\newblock Verlag f{\"u}r {W}issenschaft und {F}orschung, 2005.
\newblock (PhD thesis, 2004, Technische Universit{\"a}t Dresden, Germany).

\bibitem[BS81]{bursan81}
S.~Burris and H.P. Sankappanavar.
\newblock {\em A Course in Universal Algebra}, volume~78 of {\em Graduate Texts
  in Mathematics}.
\newblock Springer-Verlag, New York, first edition, 1981.
\newblock Corrected version available at
  http://www.thoralf.uwaterloo.ca/htdocs/ualg.html.

\bibitem[CDG{\etalchar{+}}07]{comdaugiljaclugtistom08}
H.~Comon, M.~Dauchet, R.~Gilleron, F.~Jacquemard, D.~Lugiez, S.~Tison, and
  M.~Tommasi.
\newblock Tree automata techniques and applications, 2007.

\bibitem[CDIV10]{cirdroignvog10}
M.~\'Ciri\'c, M.~Droste, J.~Ignjatovi\'c, and H.~Vogler.
\newblock Determinization of weighted finite automata over strong bimonoids.
\newblock {\em Inform. Sci.}, 180(18):3497--3520, 2010.

\bibitem[DFG16]{drofulgoe16}
M.~Droste, Z.~F{\"u}l{\"o}p, and D.~G{\"o}tze.
\newblock A {K}leene theorem for weighted tree automata over tree valuation
  monoids.
\newblock In A.-H. Dediu, J.~Janou\v{s}ek, C.~Martin-Vide, and B.~Truthe,
  editors, {\em Proceedings of Language Automata Theory and Applications (LATA
  2016)}, volume 9618 of {\em Lecture Notes in Computer Science}, pages
  452--463. Springer, Cham, 2016.

\bibitem[DFKV20]{drofulkosvog20b}
M.~Droste, Z.~F{\"u}l{\"o}p, D.~K{\'o}sz{\'o}, and H.~Vogler.
\newblock Crisp-determinization of weighted tree automata over additively
  locally finite and past-finite monotonic strong bimonoids is decidable.
\newblock In G.~Jir\'askov\'a and G.~Pighizzini, editors, {\em Descriptional
  Complexity of Formal Systems (DCFS 2020)}, volume 12442 of {\em Lecture Notes
  in Computer Science}, pages 39--51. Springer Nature Switzerland, 2020.

\bibitem[DFKV22]{drofulkosvog21}
M.~Droste, Z.~F{\"u}l{\"o}p, D.~K{\'o}sz{\'o}, and H.~Vogler.
\newblock Finite-image property of weighted tree automata over past-finite
  monotonic strong bimonoids.
\newblock {\em Theoret. Comput. Sci.}, 919:118--143, 2022.

\bibitem[DGMM11]{drogoemaemei11}
M.~Droste, D.~G{\"o}tze, S.~M{\"a}rcker, and I.~Meinecke.
\newblock Weighted tree automata over valuation monoids and their
  characterization by weighted logics.
\newblock In W.~Kuich and G.~Rahonis, editors, {\em Algebraic Foundations in
  Computer Science}, volume 7020 of {\em Lecture Notes in Computer Science},
  pages 30--55. Springer, Berlin, Heidelberg, 2011.

\bibitem[DHV15]{droheuvog15}
M.~Droste, D.~Heusel, and H.~Vogler.
\newblock Weighted unranked tree automata over tree valuation monoids and their
  characterization by weighted logics.
\newblock In A.~Maletti, editor, {\em 6th Int. Conf. on Algebraic Informatics
  (CAI 2015)}, volume 9270 of {\em Lecture Notes in Computer Science}, pages
  90--102. Springer, Cham, 2015.

\bibitem[DK21]{drokus21}
M.~Droste and D.~Kuske.
\newblock Weighted automata.
\newblock In J.-E. Pin, editor, {\em Handbook of Automata Theory}, chapter~4,
  pages 113--150. European Mathematical Society, 2021.

\bibitem[DKV09]{drokuivog09}
M.~Droste, W.~Kuich, and H.~Vogler, editors.
\newblock {\em Handbook of Weighted Automata}.
\newblock EATCS Monographs in Theoretical Computer Science. Springer-Verlag,
  2009.

\bibitem[DSV10]{drostuvog10}
M.~Droste, T.~St\"uber, and H.~Vogler.
\newblock Weighted finite automata over strong bimonoids.
\newblock {\em Inform. Sci.}, 180(1):156--166, 2010.

\bibitem[DV10]{drovog10}
M.~Droste and H.~Vogler.
\newblock Kleene and {B}{\"u}chi theorems for weighted automata and
  multi-valued logics over arbitrary bounded lattices.
\newblock In Y.~Gao, H.~Lu, S.~Seki, and S.~Yu, editors, {\em Developments in
  Language Theory (DLT 2020)}, volume 6224 of {\em Lecture Notes in Computer
  Science}, pages 160--172. Springer, 2010.

\bibitem[DV12]{drovog12}
M.~Droste and H.~Vogler.
\newblock Weighted automata and multi-valued logics over arbitrary bounded
  lattices.
\newblock {\em Theoret. Comput. Sci.}, 418:14--–36, 2012.

\bibitem[Eil74]{eil74}
S.~Eilenberg.
\newblock {\em Automata, Languages, and Machines -- Volume A}, volume~59 of
  {\em Pure and Applied Mathematics}.
\newblock Academic Press, 1974.

\bibitem[Eng75]{eng75-15}
J.~Engelfriet.
\newblock Tree automata and tree grammars.
\newblock Technical Report {DAIMI FN-10}, Inst. of Mathematics, University of
  Aarhus, Department of Computer Science, Ny Munkegade, 8000 Aarhus C, Denmark,
  1975.
\newblock ar{X}iv:1510.02036.

\bibitem[FKV21]{fulkosvog19}
Z.~F{\"u}l{\"o}p, D.~K{\'o}sz{\'o}, and H.~Vogler.
\newblock Crisp-determinization of weighted tree automata over strong
  bimonoids.
\newblock {\em Discrete Math. Theoret. Comput. Sci.}, 23(1), 2021.
\newblock $\#$18, see also arxiv.org:1912.02660v2, 2019.

\bibitem[FSV12]{fulstuvog12}
Z.~F{\"u}l{\"o}p, T.~St\"uber, and H.~Vogler.
\newblock A {B}\"uchi-like theorem for weighted tree automata over
  multioperator monoids.
\newblock {\em Theory Comput. Syst.}, 50:241--278, 2012.
\newblock Submitted on 14 October 2009, published online 28 October 2010.

\bibitem[FV09]{fulvog09new}
Z.~F{\"u}l{\"o}p and H.~Vogler.
\newblock Weighted tree automata and tree transducers.
\newblock In M.~Droste, W.~Kuich, and H.~Vogler, editors, {\em Handbook of
  Weighted Automata}, Monographs in Theoretical Computer Science. An EATCS
  Series, chapter~9, pages 313--403. Springer-Verlag, 2009.

\bibitem[FV22]{fulvog22}
Z.~F\"ul\"op and H.~Vogler.
\newblock {\em Weighted Tree Automata -- May it be a litle more?}
\newblock arXiv:2212.05529 [cs.FL], 2022.

\bibitem[FV24]{fulvog23}
Z.~F\"ul\"op and H.~Vogler.
\newblock A comparison of sets of recognizable weighted tree languages over
  specific sets of bounded lattices.
\newblock {\em International Journal of Foundations of Computer Science}, 35,
  No. 1$\&$2:51--76, 2024.

\bibitem[GFD19]{drofulgoe19}
D.~G{\"o}tze, Z.~F{\"u}l{\"o}p, and M.~Droste.
\newblock A {K}leene theorem for weighted tree automata over tree valuation
  monoids.
\newblock {\em Inf. Comput.}, 269, 2019.
\newblock Article 104445.

\bibitem[GM18]{gasmon18}
P.~Gastin and B.~Monmege.
\newblock A unifying survey on weighted logics and weighted automata.
\newblock {\em Soft Computing}, 22(4):1047--1065, 2018.

\bibitem[Gol99]{gol99}
J.S. Golan.
\newblock {\em Semirings and their Applications}.
\newblock Kluwer Academic Publishers, Dordrecht, 1999.

\bibitem[GS84]{gecste84}
F.~G{\'e}cseg and M.~Steinby.
\newblock {\em Tree Automata}.
\newblock Akad{\'e}miai Kiad{\'o}, Budapest, 1984.
\newblock ar{X}iv:1509.06233.

\bibitem[GS97]{gecste97}
F.~G{\'e}cseg and M.~Steinby.
\newblock Tree languages.
\newblock In G.~Rozenberg and A.~Salomaa, editors, {\em Handbook of Formal
  Languages}, volume~3, chapter~1, pages 1--68. Springer-Verlag, 1997.

\bibitem[HW93]{hebwei93}
U.~Hebisch and H.J. Weinert.
\newblock {\em Semirings - Algebraic Theory and Applications in Computer
  Science}.
\newblock World Scientific, 1993.

\bibitem[KS86]{kuisal86}
W.~Kuich and A.~Salomaa.
\newblock {\em Semirings, Automata, Languages}, volume~5 of {\em EATCS
  Monographs on Theoretical Computer Science EATCS Ser.}
\newblock Springer-Verlag, 1986.

\bibitem[Rad10]{rad10}
D.~Radovanovi\'c.
\newblock Weighted tree automata over strong bimonoids.
\newblock {\em Novi Sad J. Math.}, 40(3):89--108, 2010.

\bibitem[Sak09]{sak09}
J.~Sakarovitch.
\newblock {\em Elements of Automata Theory}.
\newblock Cambridge University Press, 2009.

\bibitem[Sch61]{sch61}
M.P. Sch{\"u}tzenberger.
\newblock On the definition of a family of automata.
\newblock {\em Inf. Control}, 4(2--3):245--270, 1961.

\bibitem[SS78]{salsoi78}
A.~Salomaa and M.~Soittola.
\newblock {\em Automata-Theoretic Aspects of Formal Power Series}.
\newblock Texts and Monographs in Computer Science, Springer-Verlag, 1978.

\bibitem[SVF09]{stuvogfue09}
T.~St{\"u}ber, H.~Vogler, and Z.~F{\"u}l{\"o}p.
\newblock Decomposition of weighted multioperator tree automata.
\newblock {\em Int. J. Foundations of Computer Sci.}, 20(2):221--245, 2009.

\bibitem[Wec78]{wec78}
W.~Wechler.
\newblock {\em The Concept of Fuzziness in Automata and Language Theory}.
\newblock Studien zur {A}lgebra und ihre {A}nwendungen. Akademie-Verlag Berlin,
  5. edition, 1978.

\bibitem[Wec92]{wec92}
W.~Wechler.
\newblock {\em Universal Algebra for Computer Scientists}, volume~25 of {\em
  EATCS Monographs on Theoretical Computer Science}.
\newblock Springer-Verlag, Heidelberg/Berlin, first edition, 1992.

\end{thebibliography}

\newcommand{\etalchar}[1]{$^{#1}$}

\end{document}